\documentclass[twocolumn]{aastex631}
\usepackage{float}
\usepackage{graphicx}
\usepackage{dcolumn}
\usepackage{bm}
\usepackage{amsmath}
\usepackage{hyperref}
\usepackage{mathtools}

\usepackage{xcolor}
\definecolor{pblack}{gray}{0.0}   
\definecolor{pdark}{gray}{0.55}  
\definecolor{plight}{gray}{0.8}  

\renewcommand{\d}{\mathrm{d}}
\renewcommand{\max}{\mathrm{max}}
\newcommand{\side}{\mathrm{side}}
\newcommand{\QO}{\mathrm{QO}}

\newcommand{\Commander}{\texttt{Commander}}
\newcommand{\NILC}{\texttt{NILC}}
\newcommand{\SEVEM}{\texttt{SEVEM}}
\newcommand{\SMICA}{\texttt{SMICA}}

\newcommand{\unitvec}[1]{\hat{\boldsymbol{#1}}}

\begin{document}

\title{Nearly Full-Sky Low-Multipole Cosmic Microwave Background Temperature Anisotropy:\\ III. CMB Temperature Anomalies}

\author[0000-0001-9054-1414]{Laura Herold}
\author[0000-0002-2147-2248]{Graeme E. Addison}
\author[0000-0001-8839-7206]{Charles L. Bennett}
\author[0000-0001-9694-1718]{Hayley C. Nofi}
\author[0000-0003-3017-3474]{J. L. Weiland}   

\affiliation{The William H. Miller III Department of Physics and Astronomy,\\ Johns Hopkins University, 3400 N. Charles Street, Baltimore, MD 21218 USA}

\correspondingauthor{L. Herold}
\email{lherold@jhu.edu}

\begin{abstract}
    Unexpected features have been observed in the cosmic microwave background (CMB) temperature on large scales. We revisit these CMB anomalies using new foreground-cleaned CMB temperature maps derived in a companion paper from WMAP and Planck data, which are tailored to low-resolution analysis and require only minimal masking of $1\%$ of the sky. These maps allow us to assess the impact of foreground-cleaning methods and the choice of sky cut on the significance of five commonly studied CMB anomalies. We find a notable impact of the choice of galactic mask on the significance of two anomalies: the significance of the low real-space correlation function and of the local-variance asymmetry reduces from $\sim 3\sigma$ for the Planck common mask with $26\%$ masked fraction to $\sim 2\sigma$ for the $1\%$ mask. We find good agreement between the two sky cuts for the low northern variance, $\sim 3\sigma$, and the parity asymmetry, $\sim 2\sigma$. For the quadrupole-octopole alignment, we find good agreement between the 1\%-mask result and the full-sky results in the literature, $\sim 3\sigma$. Thus using a larger fraction of the sky enabled by improved foreground cleaning reduces the significance of two commonly studied CMB anomalies. Overall, for an alternative physical model to be convincingly favored over $\Lambda$CDM with statistically isotropic Gaussian fluctuations, it would need to explain multiple CMB anomalies, or better describe some other type of measurement in addition to a CMB anomaly. 
\end{abstract}
\vspace*{5mm}


\section{Introduction}
\label{sec:intro}

A statistically isotropic and homogeneous $\Lambda$ cold dark matter ($\Lambda$CDM) model has emerged as a ``standard'' model in cosmology. While this model presents an excellent description of most data, several tensions and anomalies have been pointed out.
Parameter tensions, e.g.\ in the Hubble parameter or the parameters of a dark energy model different from $\Lambda$, may suggest new physics related to the universe's composition (see \cite{CosmoVerse:2025txj} for a recent review). In this study, we focus on features in the cosmic microwave background (CMB) known as CMB anomalies, which are present at the largest angular scales. While these features have little to no impact on the parameters of the $\Lambda$CDM model and are generally at lower significance than the parameter tensions, they seem -- taken at face value -- to challenge basic assumptions of homogeneity and isotropy.

Unexpected features in the CMB have already been pointed out in data from the Cosmic Background Explorer \citep[COBE,][]{COBE:1992syq, COBE:1992ben, Hinshaw:1996ut}, in data from the Wilkinson Microwave Anisotropy Probe \citep[WMAP,][]{WMAP:2003elm,WMAP:2008lyn, WMAP:2012fli}, and were later confirmed in data from Planck \citep{Planck:2013lks, Planck:2019evm}. The most significant features include an overall lack of power at large angular scales and thus low angular correlation; a parity asymmetry that manifests in larger power of odd compared to even multipoles of the power spectrum; an alignment of the directions defined by the quadrupole and octopole; a low variance of the temperature fluctuations on the northern hemisphere at low resolution; and a hemispherical asymmetry in the local variance \citep[for reviews see][]{Schwarz:2015cma, Abdalla:2022yfr}\footnote{Other features, which we will not cover here, are for example map-level point-parity or mirror-parity asymmetry \citep[e.g.][]{Land:2005jq}, special regions in the map like the ``cold spot'' \citep[e.g.][]{Vielva:2003et}, or spatial variations of cosmological parameters \citep{Aluri:2023xmb}.}.

Although some of these features have been reported more than 20 years ago, their origin is still unclear.
The significance of the individual CMB anomalies is of the order of a few sigma, such that individual features could be simply unlikely realizations of a universe described by the ``standard'' cosmological model. If these features are not simply due to a statistical fluke, they could be hints of new physics, improper foreground cleaning or instrument noise. The latter, however, seems unlikely since the same features were detected independently by the WMAP and Planck experiments.  

New physics to explain the CMB anomalies often requires alternative inflationary models. For example,  an inflationary phase with a low number of e-folds \citep{Powell:2006yg, Destri:2008fj, Ramirez:2012gt} or with ``causal diamonds'' \citep{Hogan:2021pap, Hogan:2023lvd} can break scale invariance and offer an explanation particularly for the low angular correlation. A coupling to super-horizon fluctuations or ``curvaton model'' can break statistical isotropy and possibly explain the hemispherical asymmetry \citep{Gordon:2005ai, Erickcek:2008jp, Erickcek:2008sm, Adhikari:2015yya, Liu:2013wfa, Kobayashi:2015qma}. Multi-field inflation, inflaton couplings and quantum gravity could explain anomalies \citep[e.g.][]{vanTent:2003mn, Wang:2013vxa, Kaiser:2013sna, Schutz:2013fua, Liu:2013iha, Braglia:2021sun, Lee:2023azx, Gaztanaga:2024vtr}. Generalized anisotropic topologies can be used to study the CMB anomalies \citep{Aurich:2013fwa, Smith:2024map}.  While these models are interesting, they often have the limitation that they only address one or two anomalies, sometimes introduce their own coincidence problem, and are often too complicated to implement numerically in order to obtain cosmological predictions.

Cosmological ``foregrounds'' like improper subtraction of the quadrupole induced by the motion of the solar system with respect to the CMB rest frame \citep{Kamionkowski:2002nd, Schwarz:2004gk, Copi:2013jna, Notari:2015kla} or temperature modulations induced by the local large-scale structure \citep{Rakic:2006tp, Inoue:2006rd, Francis:2009pt, Rassat:2013gla, Jung:2024slj} have been shown to be too small to explain the anomalies. 

The CMB anomalies could also be caused by contamination with galactic foregrounds like synchrotron emission, free-free/bremsstrahlung emission, CO line emission, spinning dust and thermal dust \citep{Slosar:2004xj, Copi:2005ff}. Due to their large size on the sky, these could naturally introduce asymmetries or modulate large-scale power.  

Although these anomaly statistics were defined a posteriori -- after they were observed in the data -- they remain a topic of active study. This is due to their appearance both in WMAP and in Planck (and some even in COBE) data. Alternative models of inflation, non-trivial topologies or models of quantum gravity are expected to leave imprints on the largest cosmological scales, which are the scales where the CMB anomalies are present \citep[for reviews see e.g.][]{Schwarz:2015cma, Luminet:2016bqv, Ashtekar:2015dja}. Our goal is not to review individual CMB anomaly theories, but to provide data analysis and statistical significance assessments of the observations, particularly as guidance for model builders.

In this work, we revisit the impact of galactic foregrounds using the new low-resolution foreground-cleaned maps published in \cite{Nofi:2025a}. These maps were obtained using CMB maps from both WMAP and Planck and foreground templates from different experiments in order to provide foreground-cleaned CMB maps at $1^\circ$ resolution, which require only minimal masking of $1\%$ of the sky. This allows us to evaluate the significance of the CMB anomalies with a larger fraction of the sky and cleaner low-resolution CMB maps than previous analyses, and thus enables us to assess the impact of the cleaning procedure and the choice of foreground mask on the significance of the anomalies. 

\cite{Nofi:2025b} found that these foreground-cleaned maps confirm the well-known low quadrupole power compared to the $\Lambda$CDM expectation at the $2.2\sigma$ level. In this work, we investigate five further commonly studied CMB anomalies.

This paper is structured as follows. We describe the data sets and methodology in Sec.~\ref{sec:methods}; we introduce the CMB anomalies studied in this work and present our results in Sec.~\ref{sec:results}; we conclude in Sec.~\ref{sec:conclusions}.


\vspace*{3mm}
\section{Data sets and methodology}
\label{sec:methods}

Notebooks and scripts to reproduce the results in this paper are publicly available\footnote{ \url{https://github.com/LauraHerold/CMBanom}}.

\subsection{Maps and masks}

We use the four new foreground-cleaned maps at $70\,\mathrm{GHz}$, $94\,\mathrm{GHz}$, $100\,\mathrm{GHz}$, and $143\,\mathrm{GHz}$ with $1^\circ$ resolution from the WMAP and Planck experiments described in \cite{Nofi:2025a}\footnote{\url{https://github.com/hnofi/ELC_Paper1}} and referred to as ``foreground-cleaned maps''. These maps were obtained by cleaning each frequency channel with six archival diffuse maps from the COBE DIRBE~\citep{Hauser:1998ri}, WMAP \citep{WMAP:2012fli}, Planck~\citep{Planck:2013fzw} experiments and the Haslam map \citep{Remazeilles:2014mba}, which are centered at frequencies dominated by CMB foregrounds including synchrotron, free-free, CO, and dust emission. The foreground maps are fit to the target maps with free overall normalization, allowing to remove foregrounds regardless of their physical emission mechanism but solely driven by their spatial morphology. This method focuses on low-resolution maps and utilizes external data sets, while previous cleaning procedures aimed at cleaning up to the highest resolution and often avoided the inclusion of external data sets \citep{Planck:2019nip}.
 
We use the $1\,\%$ mask obtained in \cite{Nofi:2025a}, which is defined by thresholding the standard deviation of the four foreground-cleaned maps such that the $1\,\%$ pixels with the largest standard deviation are masked, including a point-source mask on the Sagittarius A region. \cite{Nofi:2025a} found that the improved cleaning procedure allows to mask a fraction as small as $1\,\%$ to obtain a clean CMB map. Both maps and masks are provided at $N_\side = 128$. 

For comparison, we consider the four public Planck 2018 component-separated maps \texttt{Commander}, \texttt{NILC}, \texttt{SEVEM}, and \texttt{SMICA}\footnote{For reproducibility, we quote the explicit file names where appropriate: \texttt{COM\_CMB\_IQU-commander\_2048\_R3.00\_full.fits} for the \Commander\ map, and corresponding files for \NILC, \SEVEM, and \SMICA.} described in \cite{Planck:2018yye} and referred to as ``component-separated maps'', which are provided at an $N_\side = 2048$. Although these maps are not intended to be used without the Planck common mask applied, to facilitate comparison to the literature and to the foreground-cleaned maps, we compute the anomaly statistics for these maps also with the 1\% mask applied and on the full sky. Moreover, inpainting techniques can impact the anomaly statistics (especially for \texttt{SMICA}, \texttt{NILC} without the common mask).

Since the CMB anomalies considered in this work are present at large scales, we downgrade the resolution of the Planck 2018 component-separated maps to $N_\side = 128$, matching the resolution of the new foreground-cleaned maps. We do so using the same procedure as in \cite{Planck:2015igc}: for each map, we compute the $a_{\ell m}$ with \texttt{healpy}\footnote{\url{https://healpy.readthedocs.io}}~\citep{Gorski:2004by}, and obtain the downgraded coefficients as
\begin{equation}
    \label{eq:downgrading}
    a_{\ell m}^\mathrm{out} = \frac{b_\ell^\mathrm{out}p_\ell^\mathrm{out}}{b_\ell^\mathrm{in}p_\ell^\mathrm{in}}a_{\ell m}^\mathrm{in},
\end{equation}
where $b_\ell$ is the beam window function and $p_\ell$ the pixel window function\footnote{This uses \texttt{healpy}'s \texttt{map2alm}, \texttt{gauss\_beam}- and \texttt{pixwin}-functions.}. All maps are smoothed to a resolution of $60'$ and any monopole or dipole is removed from the full-sky maps using \texttt{healpy}'s \texttt{remove\_dipole()}.

For comparison, we also consider Planck's 2018 common mask\footnote{\texttt{COM\_Mask\_CMB-common-Mask-Int\_2048\_R3.00.fits}}, which is provided at $N_\side = 2048$ and masks about $22\,\%$ of the sky. We downgrade the mask to $N_\side = 128$ in the same way as described for the maps above, resulting in a masked fraction of 25.6\%. To recover a binary mask, we define a parent pixel as masked (pixel value 0) when a fraction of more than $10\%$ of sub-pixels are masked (i.e. below a pixel value of 0.9). For the tests of low northern variance (Sec.~\ref{sec:sigma16}), we downgrade the new foreground-cleaned and Planck-2018 component-separated maps as well as the 1\%\ and common masks to $N_\side = 16$ in the same way as described above (resulting in masked fractions of 3.6\% and 45.5\%, respectively).  The $N_\mathrm{side} = 16$ maps are smoothed to $640'$ following the convention of the Planck team  \citep[Table\ 1 in][]{Planck:2015igc}.

\subsection{Simulated maps}
\label{sec:sims}

We generate $10^5$ Gaussian statistically isotropic CMB temperature maps by drawing from the Planck-2018 bestfit $\Lambda$CDM model\footnote{\texttt{COM\_PowerSpect\_CMB-base-plikHM-TTTEEE-lowl-lowE-lensing-\\ minimum-theory\_R3.01.txt}}~\citep{Planck:2018vyg} using \texttt{healpy}'s \texttt{synfast}-function. The maps are generated at an $N_\side = 128$ to match the foreground-cleaned maps' resolution. All maps are stored for reproducibility and the same set of maps is used for all tests. For the tests of hemispherical asymmetry, we downgrade to $N_\side = 16$ in the same way as described above. 

Since the cleaning procedure in \cite{Nofi:2025a} is based on the morphological shape of the foregrounds, there could be a chance correlation of the morphology of the CMB and the foregrounds, leading to possible over- or under-subtraction of foreground templates. As described in \cite{Nofi:2025b}, we study the impact of such a CMB--foreground chance correlation by adding the foreground templates multiplied by the coefficients obtained in \cite{Nofi:2025a} to $10^4$ simulated CMB maps and cleaning these simulations with the same procedure as the real CMB maps. We report the shifts in probability to exceed (PTE or $p$-value) when using the CMB-only simulations and cleaned-CMB simulations in the respective subsections. Since the shifts are small or do not impact the conclusions, our default PTE values use the $10^5$ CMB-only simulations.

\subsection{Power spectra and correlation function}

The CMB temperature fluctuations are commonly expanded in terms of spherical harmonics, $Y_{\ell m}(\theta, \phi)$,
\begin{equation}
    \label{eq:sph_harm_exp}
    T(\theta, \phi) = \sum_{\ell=1}^\infty \sum_{m=-\ell}^\ell a_{\ell m}\, Y_{\ell m}(\theta, \phi) ,
\end{equation}
where all cosmological information is contained in the coefficients $a_{\ell m}$. The power spectrum is then given by the two-point function $\langle a_{\ell m}a^*_{\ell' m'}\rangle = C_\ell\, \delta_{\ell\ell'}\, \delta_{m m'}$. A common estimator for the power spectra is the ``pseudo-$C_\ell$'' estimator, 
\begin{equation}
    \hat{C}_\ell = \frac{1}{2\ell+1} \sum_{m=-\ell}^{+\ell} |\hat{a}_{\ell m}|^2,
\end{equation}
which is then corrected for the effect of the mask or weighting to give an unbiased estimate of the true power spectrum. The real-space correlation function of the temperature fluctuations is given by
\begin{equation}
    \label{eq:C_theta}
    C(\theta) = \langle T(\unitvec n_1) T(\unitvec n_2)\rangle = \sum_{\ell=2}^\infty \frac{2\ell+1}{4\pi} C_\ell P_\ell (\cos\theta),
\end{equation}
which can be related to the power spectrum, $C_\ell$, via the Legendre polynomials $P_\ell$, where the second equality holds only for the full sky. The angle $\theta$ is the angle between the two unit vectors, $\unitvec n_1$, $\unitvec n_2$, on the sky, i.e.\ $\cos\theta = \unitvec n_1 \cdot \unitvec n_2$.

\begin{table*}[ht]
     \hspace*{-1cm}
     \begin{tabular}{ll|cccc|cccc}
     \multicolumn{2}{c}{$p$-values [\%]} &70\,GHz &94\,GHz &100\,GHz &143\,GHz &\Commander &\NILC &\SEVEM &\SMICA\\
     \hline
            &full sky: &\textcolor{plight}{8.35} &\textcolor{plight}{8.91} &\textcolor{plight}{7.40} &\textcolor{plight}{6.94} &\textcolor{pdark}{4.62} &\textcolor{plight}{6.65} &\textcolor{pdark}{4.02} &\textcolor{plight}{6.20} \\
    $S_{1/2}$       &1\% mask: &\textcolor{plight}{7.30} &\textcolor{plight}{7.14} &\textcolor{plight}{6.28} &\textcolor{plight}{5.78}          &\textcolor{plight}{5.15} &\textcolor{plight}{5.94} &\textcolor{pdark}{3.02} &\textcolor{plight}{5.88} \\
            &com. mask: &\textcolor{pblack}{0.19} &\textcolor{pblack}{0.24} &\textcolor{pblack}{0.20} &\textcolor{pblack}{0.19} &\textcolor{pblack}{0.15} &\textcolor{pblack}{0.13} &\textcolor{pblack}{0.14} &\textcolor{pblack}{0.12} \\
    \hline
        &full sky: &\textcolor{pdark}{3.36} &\textcolor{pdark}{2.94} &\textcolor{pdark}{3.67} &\textcolor{pdark}{4.04} &\textcolor{plight}{5.36} &\textcolor{pdark}{2.81} &\textcolor{plight}{12.77} &\textcolor{pdark}{3.63} \\
    $R^{27}$        &1\% mask: &\textcolor{pdark}{2.84} &\textcolor{pdark}{3.34} &\textcolor{pdark}{3.06} &\textcolor{pdark}{3.27}      &\textcolor{pdark}{2.96} &\textcolor{pdark}{2.43} &\textcolor{plight}{6.95} &\textcolor{pdark}{2.64} \\
        &com. mask: &\textcolor{plight}{5.36} &\textcolor{plight}{5.85} &\textcolor{plight}{5.08} &\textcolor{plight}{5.29} &\textcolor{plight}{5.83} &\textcolor{plight}{5.68} &\textcolor{plight}{6.01} &\textcolor{plight}{5.96} \\
    \hline
        &full sky: &\textcolor{pblack}{0.11} &\textcolor{pblack}{0.07} &\textcolor{pblack}{0.06} &\textcolor{pblack}{0.05} &\textcolor{pdark}{0.46} &\textcolor{pblack}{0.04} &\textcolor{pdark}{1.35} &\textcolor{pblack}{0.10} \\
    $S_\mathrm{QO}$ &1\% mask: &\textcolor{pblack}{0.13} &\textcolor{pblack}{0.10} &\textcolor{pblack}{0.08} &\textcolor{pblack}{0.06} &\textcolor{pdark}{0.63} &\textcolor{pblack}{0.06} &\textcolor{pblack}{0.13} &\textcolor{pblack}{0.11} \\
        &com. mask: &\textcolor{plight}{10.18} &\textcolor{plight}{12.26} &\textcolor{plight}{8.83} &\textcolor{plight}{7.98} &\textcolor{plight}{7.31} &\textcolor{plight}{9.00} &\textcolor{plight}{14.19} &\textcolor{plight}{6.81} \\
    \hline
        &full sky: &\textcolor{pdark}{0.59} &\textcolor{pdark}{0.68} &\textcolor{pdark}{0.60} &\textcolor{pdark}{0.65} &\textcolor{pdark}{0.57} &\textcolor{pdark}{0.51} &\textcolor{pdark}{1.94} &\textcolor{pdark}{0.48} \\
    $\sigma_{16}^2$ &3.6\% mask: &\textcolor{pdark}{0.35} &\textcolor{pdark}{0.40} &\textcolor{pdark}{0.35} &\textcolor{pdark}{0.34} &\textcolor{pdark}{0.29} &\textcolor{pdark}{0.40} &\textcolor{pblack}{0.19} &\textcolor{pdark}{0.37} \\
        &com. mask: &\textcolor{pblack}{0.14} &\textcolor{pblack}{0.16} &\textcolor{pblack}{0.17} &\textcolor{pblack}{0.15} &\textcolor{pdark}{0.32} &\textcolor{pdark}{0.36} &\textcolor{pblack}{0.13} &\textcolor{pdark}{0.36} \\
    \hline
        &full sky: &\textcolor{pdark}{3.99} &\textcolor{pblack}{0.00} &\textcolor{pdark}{4.13} &\textcolor{plight}{8.62} &\textcolor{pblack}{0.00} &\textcolor{pdark}{0.86} &\textcolor{pblack}{0.00} &\textcolor{pdark}{0.77} \\
    $A_\mathrm{LV}$ &1\% mask: &\textcolor{pdark}{2.42} &\textcolor{pdark}{2.26} &\textcolor{pdark}{2.18} &\textcolor{pdark}{2.77} &\textcolor{pblack}{0.29} &\textcolor{pdark}{0.71} &\textcolor{pblack}{0.00} &\textcolor{pdark}{0.78} \\
        &com. mask: &\textcolor{pblack}{0.16} &\textcolor{pblack}{0.14} &\textcolor{pblack}{0.13} &\textcolor{pblack}{0.13} &\textcolor{pblack}{0.11} &\textcolor{pblack}{0.10} &\textcolor{pblack}{0.10} &\textcolor{pblack}{0.10} \\
    \hline
    \end{tabular}
    \caption{Significance of the anomaly tests in terms of PTE ($p$-values) in percent of the five statistics considered in this work: Low correlation, $S_{1/2}$; parity asymmetry, $R^{27}$; quadrupole-octopole alignment, $S_\mathrm{QO}$; low northern variance at $N_\side=16$, $\sigma_{16}^2$; and hemispherical local-variance asymmetry, $A_\mathrm{LV}$. The 1\% mask covers $1.0\%$ of the sky at $N_\side = 128$ ($3.6\%$ at $N_\side = 16$ used for $\sigma^2_{16}$); the Planck common mask covers $25.6\%$ of the sky at $N_\side=128$ ($45.5\%$ at $N_\side = 16$). As a visual guide, we display $p < 0.27\%$ ($>3\sigma$) in black, $0.27\% <p < 4.55 \%$ (between $2 \sigma$ and $3\sigma$) in gray and $p > 4.55 \%$ ($<2\sigma$) in light gray.}
    \label{tab:p-values}
\end{table*}

We use \texttt{Polspice}\footnote{\url{https://www2.iap.fr/users/hivon/software/PolSpice/}} \citep{Szapudi:2000xj, Chon:2003gx, Challinor:2011} to obtain the mask-deconvolved $C_\ell$ and $C(\theta)$ of the real and simulated maps. \texttt{Polspice} corrects for mode-coupling induced by the galactic mask with a mode-coupling matrix method \citep{Chon:2003gx, Hivon:2001jp}. To facilitate comparison to previous works, which are based on auto power spectra and auto correlation functions, we base our tests on auto spectra and functions as well. The mask-deconvolved $C_\ell$ obtained with \texttt{Polspice} are corrected for the beam and pixel window function by weighting with $b_\ell^\mathrm{out}$ and $p_\ell^\mathrm{out}$ as defined in Eq.~\ref{eq:downgrading}.

For comparison, we also consider the quadratic maximum likelihood (QML) estimate\footnote{\texttt{COM\_PowerSpect\_CMB-TT-full\_R3.01.txt}} of the $C_\ell$ published by the Planck team where applicable. The QML-$C_\ell$ are extracted from the \Commander\ map using a Blackwell-Rao estimator \citep{Chu:2004zp, Rudjord:2008vc}, which is an unbiased estimator with smaller variance than the pseudo-$C_\ell$ estimator.


\section{Results: CMB anomaly tests}
\label{sec:results}

In this section, we define the five anomaly tests considered in this work and present our results. We quote significances in terms of PTE and in terms of $\sigma$ corresponding to a two-sided test and Gaussian distribution, i.e. $p = 31.73 \%$, $p = 4.55 \%$ and $p = 0.27\%$ corresponds to $1\sigma$, $2\sigma$, and $3\sigma$, respectively. We refer to significances $<3\sigma$ as ``mild'', and significances $[3\sigma,\, 4\sigma]$ as ``moderate''.  A summary of the anomaly statistics and $p$-values are given in Table~\ref{tab:p-values} and Fig.~\ref{fig:anomaly_stats_all}. We quote the absolute values of the anomaly statistics in App.~\ref{app:all}.


\subsection{Low correlation, $S_\mu$}
\label{sec:S_12}

\begin{figure}[t]
    \centering
    \includegraphics[width=1\linewidth]{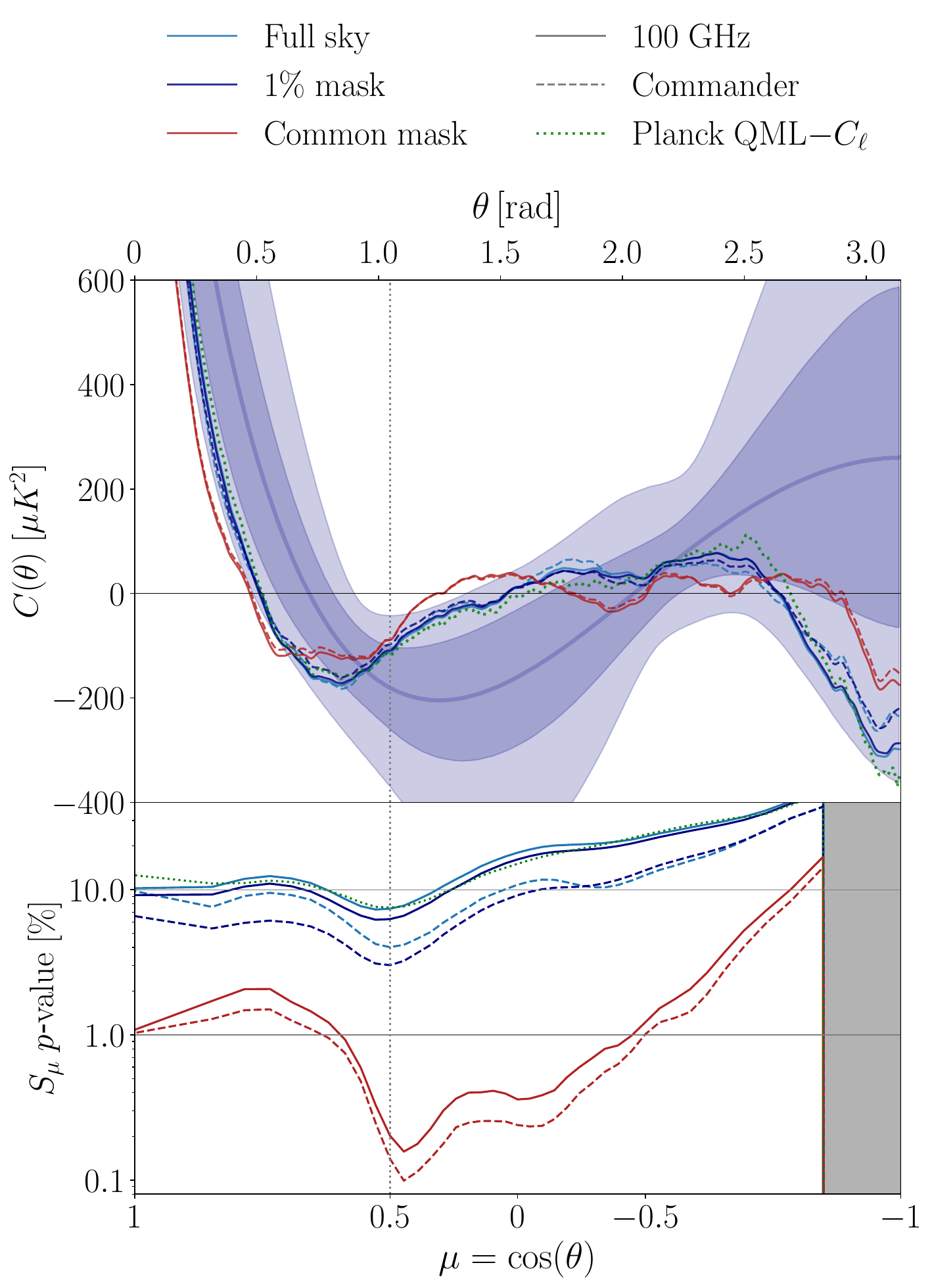}
    \caption{\textit{Top: } Angular correlation function, $C(\theta)$, for the $100\,\mathrm{GHz}$ cleaned map and \Commander{} map using the full sky (light blue), 1\% mask (dark blue), and Planck common mask (red). For comparison, we show $C(\theta)$ computed from the public QML-$C_\ell$ (green dotted). When applying the common mask, $C(\theta)$ is closer to zero for $\theta>60^\circ$ (vertical dotted line) than expected from the $\Lambda$CDM simulations ($1\,\sigma$ and $2\,\sigma$ shaded band, applying the 1\% mask); less so for all other sky configurations. \textit{Bottom: } Significance in $p$-values of the $S_\mu$-statistic, Eq.~\eqref{eq:S_mu}, as a function of the integration limit $\mu = \cos(\theta)$. Moderately low PTE values, $p<1\%$, are obtained for the common mask and $-0.5 \lesssim \mu \lesssim 0.7$.}
    \label{fig:corr}
\end{figure}

The real-space angular correlation function, Eq.~\eqref{eq:C_theta}, was observed to be closer to zero than expected at large angular scales in COBE, WMAP and Planck data \citep{Hinshaw:1996ut, WMAP:2003ivt,Planck:2013lks}. 
The correlation function is a sensitive probe of the largest angular scales, which is the regime where alternative models of inflation, non-trivial topologies and quantum gravity are expected to leave imprints. Compared to the power spectrum, it is more sensitive to features localized in real space.
We show the angular correlation function, $C(\theta)$, of the foreground-cleaned $100\,\mathrm{GHz}$ map along with the \Commander{} map in Fig.~\ref{fig:corr}, which are representative for the other foreground-cleaned and component-separated maps (all maps are shown in App.~\ref{app:all}). We confirm that $C(\theta)$ appears closer to zero than expected from Gaussian-isotropic $\Lambda$CDM simulations (blue shaded $1\sigma$ and $2\sigma$ regions) when applying the Planck common mask (red). Using the $1\,\%$ confidence mask, we obtain a $C(\theta)$ that closely resembles the full-sky result (light blue), both of which visually show a larger deviation from zero than the common-mask result. The $100\,\mathrm{GHz}$ and \Commander{} $C(\theta)$ agree well, which indicates that the impact of common mask compared to the 1\%-mask and full-sky result dominates over the impact of the cleaning procedure. We observe small differences between $100\,\mathrm{GHz}$ and \Commander{} on the largest scales caused by the lower quadrupole in the foreground-cleaned maps \citep[see][]{Nofi:2025b}. For comparison, we show the correlation function computed from the public Planck{} QML $C_\ell$, which qualitatively agrees with the 1\%-mask and full-sky results. 

\begin{figure*}[h]
    \includegraphics[width=1\textwidth]{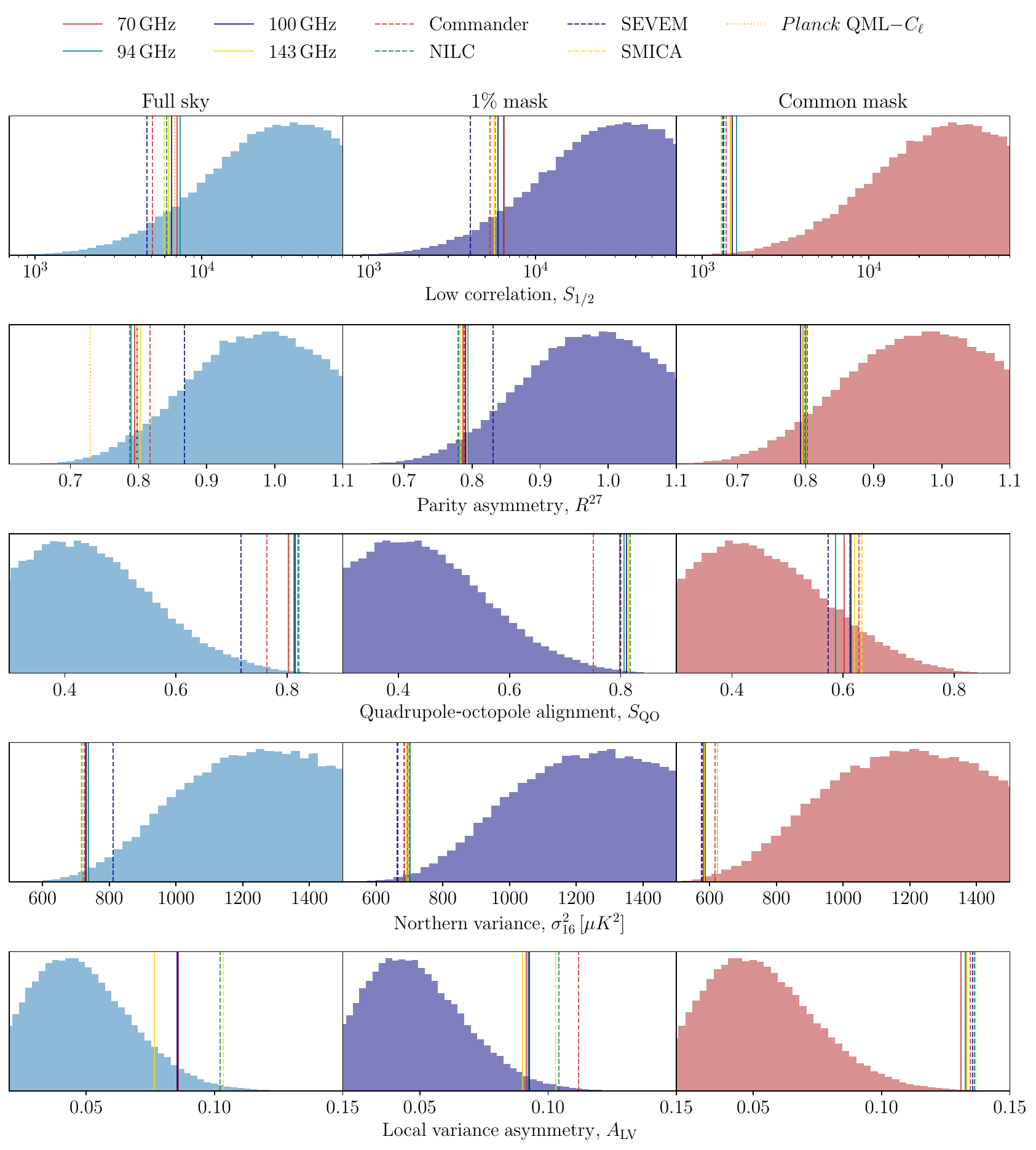}
    \caption{Estimators of the five CMB anomalies considered in this work (rows) applied to the four foreground-cleaned maps \citep[][$70\, \mathrm{GHz}$ - $143\, \mathrm{GHz}$]{Nofi:2025a} and the component-separated maps \citep[][\Commander{} - \SMICA]{Planck:2018yye} for three sky cuts (columns). The histograms show the distributions obtained from $10^5$ statistically isotropic Gaussian $\Lambda$CDM simulations.}
    \label{fig:anomaly_stats_all}
\end{figure*}

To quantify the significance of the low angular correlation, it is common to use the statistic \citep{WMAP:2003elm}:
\begin{equation}
    \label{eq:S_mu}
    S_\mu = \int_{-1}^\mu[C(\theta)]^2\, \d(\cos\theta),
\end{equation}
where typically $\mu = \cos\theta = 1/2$. Hence, $S_{1/2}$ is sensitive to the deviation of $C(\theta)$ from zero between $\theta = 60^\circ$ and $\theta = 180^\circ$. Since, in practice, our data are binned in bins of $\theta$, we approximate Eq.~\eqref{eq:S_mu} as a sum:
\begin{equation}
    S_\mu = \sum_i^{\{\cos\theta_i<\mu\}} [C(\theta_i)]^2\cdot \delta\theta_i,
\end{equation}
which allows for faster evaluation. We verified that this approximation agrees well with interpolating and integrating the binned $C(\theta)$ or using the recursion relation described in \cite{Copi:2008hw, Muir:2018hjv}. Note that we ensure consistency by treating the simulations in exactly the same way as the real maps.

We show the $S_{1/2}$-statistic for the eight different maps and three different sky cuts in the top row of Fig.~\ref{fig:anomaly_stats_all}. For a given masking choice, we find good agreement for all \cite{Nofi:2025a} foreground-cleaned maps ($70\, \mathrm{GHz}$ - $143\, \mathrm{GHz}$), which yield $S_{1/2}$-values slightly closer to the mean of the $10^5$ Gaussian statistically isotropic $\Lambda$CDM simulations (histograms, note the $\log$ scale on the $x$-axis) than the \cite{Planck:2018yye} component-separated maps (\Commander{} - \SMICA). We find that the $S_{1/2}$-statistic of the maps with the Planck common mask applied is in the tail of the histogram with a moderately significant PTE of $p = 0.19 \% - 0.24\%$ for the foreground-cleaned maps (see Table~\ref{tab:p-values}, corresponding to $3.0\sigma-3.1\sigma$) and $p=0.12\% - 0.15\%$ (corresponding to $3.2\sigma$) for the component-separated maps. While the Planck common mask is usually applied in the literature when measuring $S_{1/2}$, the improved foreground cleaning allows us to explore this statistic with a larger sky fraction. When applying the 1\% mask, we find $S_{1/2}$-values closer to the mean of the simulations and a notably reduced significance of this statistic with $p = 5.8\% - 7.3\%$ for the foreground-cleaned maps ($1.8\sigma-1.9\sigma$). Although the component-separated maps are not intended for use without the common mask, for comparison, we also quote the results of these maps with the 1\% mask applied, finding $p = 3.0\% - 5.9\%$ ($1.9\sigma - 2.2\sigma$) in good agreement with the foreground-cleaned maps. Even though none of the maps are expected to be free of foreground contamination on the full sky, for comparison to previous work we quote the anomaly statistics also for all maps without any galactic mask applied. For the full sky, we find $p= 6.9\% - 8.9 \%$ ($1.7\sigma - 1.8\sigma$) for the foreground-cleaned and $p=4.0\% - 6.7\%$ ($1.8\sigma - 2.1\sigma$) for the component-separated maps.

As described in Sec.~\ref{sec:sims} and \cite{Nofi:2025b}, in order to probe the impact of a possible chance correlation between the CMB and the foreground templates applied in the cleaning procedure, we compare to $10^4$ CMB simulations that were cleaned with the same methodology as the real maps. Computing the significance of $S_{1/2}$ with the cleaned-CMB simulations compared to the CMB-only simulations results in shifts of $\Delta p = |p_\mathrm{CMB-only} - p_\mathrm{cleaned-CMB}| \leq 1.1\%$ for full sky, $\Delta p \leq 0.9\% $ for the 1\%, and $\Delta p \leq 0.05\% $ for the common mask. Since these shifts do not impact the conclusions, we quote the significances using the $10^5$ CMB-only simulations.

For comparison, we consider the $S_{1/2}$-statistic of the public Planck QML-$C_\ell$. Although the QML-$C_\ell$ were derived using a different mask -- making direct comparison with our simulations difficult -- we find good agreement of the histograms of the simulations across the various sky cuts considered in this work ($0\%$, $1\%$, and $26\%$ masked fraction). Thus we show the $S_{1/2}$-values of the QML-$C_\ell$ for comparison in the full-sky panel of Fig.~\ref{fig:anomaly_stats_all}, which are in good agreement with the other maps for the full-sky or 1\%-mask case, albeit less significant than the common-mask case.

We explore the significance of the $S_\mu$-statistic for varying integration limit, $\mu=\cos(\theta)$ in Eq.~\eqref{eq:S_mu}, in Fig.~\ref{fig:corr} for the $100\, \mathrm{GHz}$ and \Commander{} maps (representative for the other maps shown in App.~\ref{app:all}). For all $\mu$, the 1\%-mask and full-sky results are less significant with $p\gtrsim 3\%$ (corresponding to $\lesssim 2\sigma$) than the common-mask result. For the common mask, we find PTE with $p\lesssim 1\%$ (corresponding to $>2.6\sigma$) for $-0.5 \lesssim \mu \lesssim 0.7$. We find a similar behavior for all maps considered in this work (shown in App.~\ref{app:all}). Thus there is a dependence of $S_\mu$ on the integration limit $\mu$, which could be included in the computation of the significance via a look-elsewhere correction (see discussion in Sec.~\ref{sec:conclusions}).

The low $p$-values of the $S_{1/2}$-statistic using the common mask are in good agreement with previous results in the literature \citep{Planck:2019evm, Jones:2023ncn}, although our values are slightly more significant than those quoted in \cite{Muir:2018hjv}, which might be due to the use of the Planck 2015 component-separated maps in their analysis (vs.\ 2018 maps here). Our findings of a significant impact of the choice of masking aligns with some previous works in the literature. \cite{Pontzen:2010yw} argued that the alignment of the quadrupole, $C_2$, and octopole, $C_3$ (see Sec.~\ref{sec:SQO}) lead to more power in the region of the galactic plane, which is largely covered by the common mask; this ``masking'' of $C_2$, $C_3$ and thus lower power in $C_2$, $C_3$ can result in a lower $C(\theta)$. \cite{Gruppuso:2013dba} illustrated that using larger galactic masks leads to more significant $p$-values of $S_{1/2}$. Moreover, our results using the 1\% mask are consistent with the ones inferred from the public MLE-$C_\ell$ \citep[c.f. discussion in][]{Efstathiou:2009di}.

Thus with the new \cite{Nofi:2025a} foreground-cleaned maps, which allow us to confidently use a larger fraction of the sky with only $1\%$ masking, we find that the correlation function deviates less significantly from the Gaussian statistically isotropic $\Lambda$CDM expectation as measured by the $S_{1/2}$-statistic (at the level of $\sim 2\sigma$), than when applying the Planck common mask ($\sim 3\sigma$).
\vspace{4mm}


\subsection{Parity asymmetry, $R^{\ell_\max}$}
\label{sec:R27}

A possible parity asymmetry of the CMB can be studied by investigating whether the CMB sky is invariant with respect to point reflections around the origin, $\hat{\boldsymbol{x}}' =  -\hat{\boldsymbol{x}}$ \citep{Land:2005jq}. Due to the symmetry of the spherical harmonics, even (odd) multipoles, $C_\ell$, contribute only to the even (odd) maps, $M^\pm = \frac{1}{2} (M\pm M')$. Thus the power in even compared to odd $C_\ell$ is a measure of the (a-)symmetry of the CMB under point reflections. Parity is assumed to be a fundamental symmetry of the universe. A violation of point-parity asymmetry would violate statistical homogeneity since it would require the observer to be located in the focal point of the symmetry and leave different locations in the Universe distinguishable. Galactic foregrounds, on the other hand, naturally lead to a parity asymmetry since they are located in the Galactic plane. Thus parity asymmetry can hint at a residual contamination with foregrounds \citep[e.g.][]{Schwarz:2015cma}.

\begin{figure}
    \centering
    \includegraphics[width=1\linewidth]{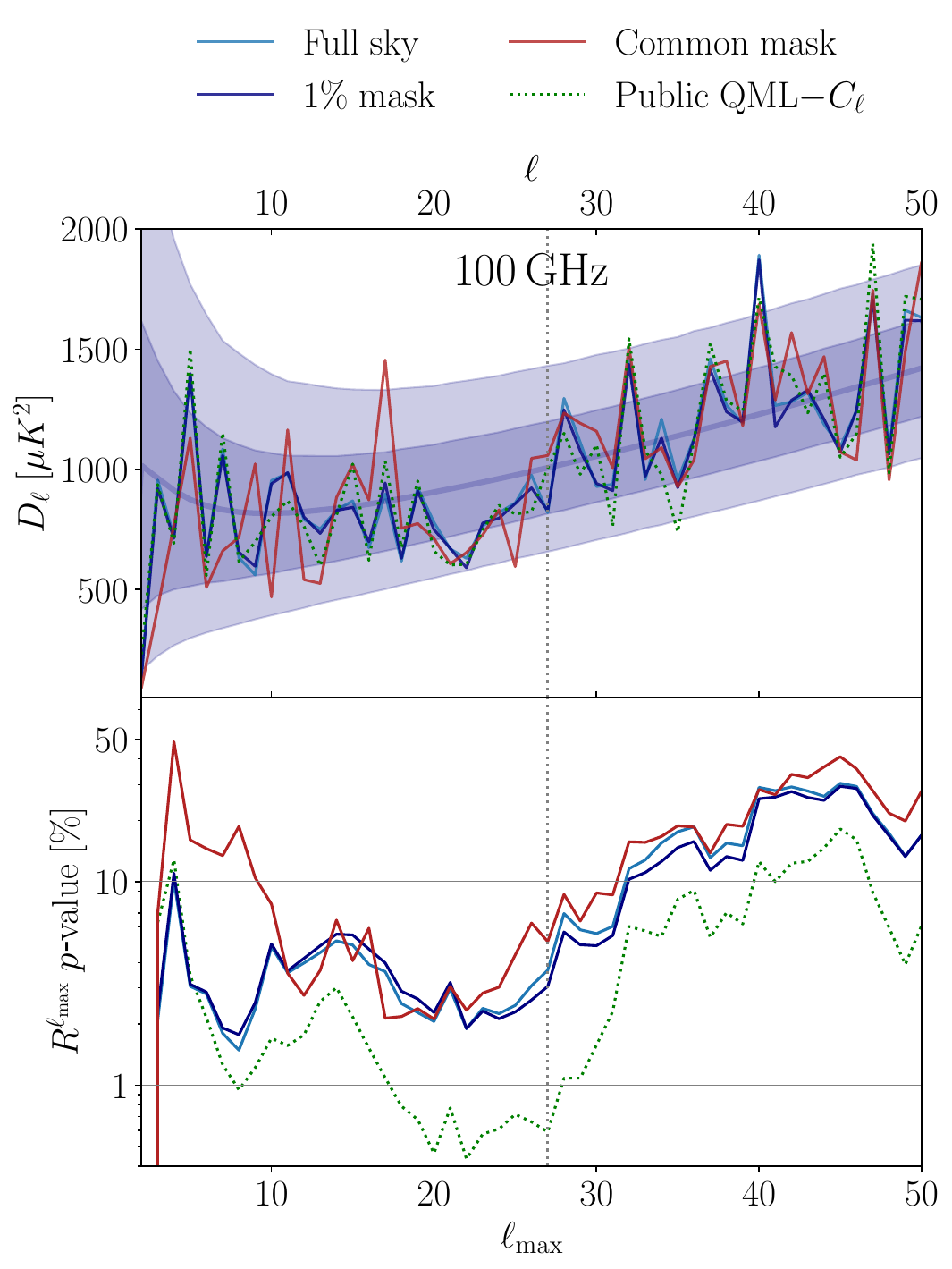}
    \caption{\textit{Top:} CMB power spectra, $D_\ell$, of the $100\, \mathrm{GHz}$ map (solid) at large angular scales for three sky cuts as indicated in the legend. For comparison, we show the public Planck{} QML-$D_\ell$ (dotted green). The ``parity asymmetry'' manifests itself as larger power for odd compared to even $\ell$. \textit{Bottom:} $p$-values of $R^{\ell_\max}$ as a function of $\ell_\max$ for the same sky cuts as above. The baseline choice $\ell_\max = 27$ (vertical dotted line) gives more significant $p$-values than higher values of $\ell_\max$; the Planck{} QML-$D_\ell$ exhibit lower $p$-values than the $100\, \mathrm{GHz}$ map.}
    \label{fig:R_lmax}
\end{figure}

We first show the power spectra, $D_\ell = \ell (\ell + 1)/(2\pi)\, C_\ell$ for $\ell \le 50$, of the foreground-cleaned $100\,\mathrm{GHz}$ map in the top panel of Fig.~\ref{fig:R_lmax} for three sky cuts. The $D_\ell$ for the full sky and 1\% mask show close agreement, while the public QML-$D_\ell$ and the common-mask $D_\ell$ show small differences compared to these.

A common measure of parity asymmetry is to compare the power in even and odd temperature multipoles \citep{Kim:2010gf}:
\begin{equation}
    \label{eq:R_ell}
    R^{\ell_\max} = \frac{D_+(\ell_\max)}{D_-(\ell_\max)},
\end{equation}
with
\begin{equation}
    D_{+,-} = \frac{1}{\ell_\mathrm{tot}^{+, -}} \sum_{\ell = 2(+,-)}^{\ell_\max} \frac{\ell(\ell+1)}{2\pi} C_\ell,
\end{equation}
where the $+$ ($-$) indicates a sum over even (odd) multipoles. An excess in odd compared to even multipoles was reported for WMAP \citep{Land:2005jq, Kim:2010gd, Gruppuso:2010nd} and Planck \citep{Planck:2015igc} at the $2\sigma-3\sigma$ level for $\ell_\max \in [20,\, 30]$. Since $\ell_\max = 27$ is one of the most common choices due to the comparatively high significance of $R^{27}$, we adopt this choice for our baseline analysis, but explore the dependence on $\ell_\max$ later. 

We show the $R^{27}$-statistic for the eight maps and three sky cuts in the second row of Fig.~\ref{fig:anomaly_stats_all}. We find good agreement for all maps with PTE $p=5.1\% - 6.0\%$ for the Planck common mask, $p = 2.6\% - 3.3\%$ for the 1\% mask, and $p = 2.8\% - 4.0\%$ for the full sky. We excluded the \Commander{} and \SEVEM{} maps from the 1\%-mask and full-sky cases, since these represent outliers with even larger PTE, which is possibly caused by residual foreground contamination. Thus for all cleaning procedures and sky cuts we find only a mild significance $<2.3\sigma$.

We test the impact of the CMB-foreground chance correlation by computing the PTE with simulations that include mock foreground cleaning. We find small shifts in the PTE of $\Delta p \leq 0.3\%$ for the full sky and 1\% mask and $\Delta p \leq 0.1\%$ for the common mask. Thus we use the $10^5$ CMB-only simulations to quote significances of $R^{\ell_\max}$.

Since we find good agreement between the histograms of the CMB simulations for the different sky cuts, for comparison we show the Planck QML-$C_\ell$ result in the full-sky panel of Fig.~\ref{fig:anomaly_stats_all}, although a direct comparison would require re-computing the CMB simulations with the QML mask applied. The QML-$C_\ell$ are farther from the mean of the simulations (corresponding to $p=0.6\%$, or about $2.7\sigma$ when computing the PTE with the full-sky simulations).

In the bottom panel of Fig.~\ref{fig:R_lmax}, we explore the impact of the choice of $\ell_\max$ on the $R^{\ell_\max}$-statistic for the $100\,\mathrm{GHz}$ cleaned map representative for all maps, which are shown in App.~\ref{app:all}. For the 1\% mask, PTE values between $2\% \lesssim p \lesssim 5\%$ ($2\sigma-2.3\sigma$) can be found in the range $\ell_\max = 5 - 30$, while for $\ell_\max >30$ the $p$-values increase. 
The QML-$C_\ell$ show moderately significant PTE below $p<1\%$ ($>2.6\sigma$) for $\ell = 18 - 27$. While the significance of the $R^{\ell_\max}$-statistic is low, there is an additional dependence of $R^{\ell_\max}$ on $\ell_\mathrm{max}$. In principle, this dependence motivates a look-elsewhere correction, which would further decrease the significance of this statistic.

Our results for the \cite{Planck:2018vyg} component-separated maps are consistent with previous studies \citep{Jones:2023ncn, Muir:2018hjv, Planck:2019evm}. 

In summary, for the $R^{27}$-statistic we find PTE values in good agreement for all maps and all sky cuts considered in this work, which correspond to a Gaussian significance of about $2\,\sigma$. Thus we confirm the mild significance of this feature found in previous studies with the new cleaned CMB maps using a larger fraction of the sky.


\subsection{Alignments of multipoles, $S_{QO}$}
\label{sec:SQO}

The quadrupole and octopole of the CMB temperature map ($\ell = 2, 3$) was pointed out to be unexpectedly planar and aligned in both WMAP and Planck data \citep{deOliveira-Costa:2003utu, Planck:2013lks}. The distribution of the directions of the CMB multipoles is a probe of the statistical isotropy posited by the cosmological principle. Due to their possible sensitivity to inflationary physics, topology and quantum gravity, the lowest multipoles have received particular attention.
There are several approaches to define a direction of a multipole; here we follow the approach suggested in \cite{Copi:2003kt} using ``Maxwell's multipole vectors''. Instead of expanding the $\ell$-th multipole $T_\ell$ in terms of spherical harmonics $Y_{\ell m}$, Eq.~\ref{eq:sph_harm_exp}, one can expand it in terms of $\ell$ multipole vectors (MVs), $\hat v_\ell$, and one amplitude, $\lambda_\ell$,
\begin{equation}
    \label{eq:MVs}
    T_\ell (\theta, \phi)  =  \lambda_\ell \nabla_{\hat v_2} \dots \nabla_{\hat v_\ell}\ \frac{1}{r}\ \Bigg|_{r=1},
\end{equation}
where $\nabla_{\hat v} = \hat v \cdot \vec \nabla$. The MVs are head-less unit vectors, i.e.\ the overall sign is not defined. Eq.~\eqref{eq:MVs} ensures that for a given direction $\hat{n} = (\theta, \phi)$, projecting $\hat{n}$ onto the axes spanned by the MVs gives $T_\ell (\hat{n})/\lambda_\ell$. Since the MVs are computed directly from the $a_{\ell m}$, they are only well defined on the full sky. Nevertheless, in the following we will show results obtained using the 1\% and common masks, but we caution that these have been obtained from cut-sky $a_{\ell m}$. Due to the smaller masked fraction of the 1\% mask compared to the Planck common mask ($26\%$), we expect the 1\% mask to be closer to the full-sky result.

To probe orthogonality of MVs, it is convenient to define the oriented-area vectors \citep[OAVs,][]{Copi:2003kt}
\begin{equation}
    \vec w^{(\ell, i, j)} = \pm (\hat{v}^{(\ell,i)} \times \hat v^{(\ell, j)}),
\end{equation}
where $i,\, j \in (1, \dots, \ell)$. The OAVs define the vector orthogonal to the plane spanned by two MVs of a given $\ell$, where the length of the OAV corresponds to the area of the parallelogram spanned by the respective MVs.

\begin{figure}[t]
    \centering
    \includegraphics[width=1\linewidth]{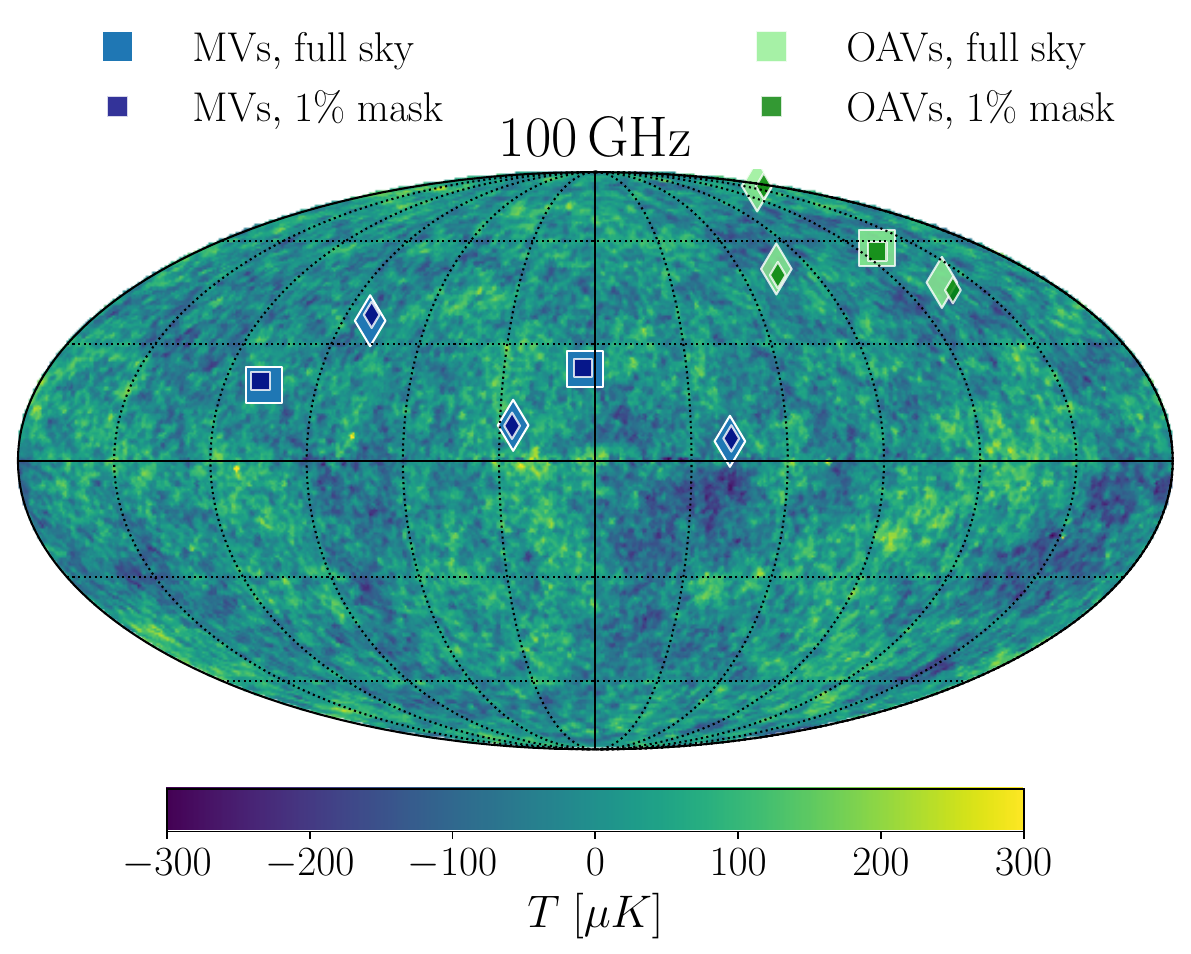}
    \caption{Multipole vectors (MVs) and oriented area vectors (OAVs) of the quadrupole (squares) and octopole (diamonds) obtained from the $100\,\mathrm{GHz}$ foreground-cleaned map (background) for the full sky (lighter colors) and 1\% mask (darker colors); the common mask is omitted (see text). The quadrupole spans one plane (two MVs) defined by one OAV, the octopole spans three planes (three MVs, three OAVs). The planes of the quadrupole and octopole MVs are approximately aligned, i.e.\ their OAVs point to a similar direction on the sky.}
    \label{fig:alignment}
\end{figure}

We compute the MVs and OAVs using \texttt{polyMV}\footnote{\url{https://github.com/oliveirara/polyMV}} \citep{Oliveira:2018sef}. The MVs and OAVs obtained from the foreground-cleaned $100\,\mathrm{GHz}$ map are shown in Fig.~\ref{fig:alignment} for the full sky and 1\% mask (all other maps are shown in App.~\ref{app:all}). The quadrupole is fully characterized by two MVs and one OAV, while the octopole is characterized by three MVs and three OAVs. The quadrupole-octupole alignment is characterized by the quadrupole's and octopole's MVs lying in a plane, which leads to OAVs pointing to a similar direction on the sky. To avoid over-crowding of the figure, the common-mask vectors are not shown; the MVs with the common mask applied show less alignment in a plane and thus less parallel OAVs. 

To probe the significance of the alignment of the quadrupole and octopole, it is common to use the simple statistic \citep{Copi:2003kt}: 
\begin{equation}
    S_\QO = \frac{1}{3} \sum_{i=1}^{2}\sum_{j=i+1}^3 |\vec w^{(2;1,2)}\cdot \vec w^{(3;i,j)}|,
\end{equation}
which quantifies the degree of alignment of the single quadrupole OAV with the three octopole OAVs by means of a scalar product. 

We show the $S_\mathrm{QO}$-statistic for the eight maps and three sky cuts in the third row of Fig.~\ref{fig:anomaly_stats_all}. We find comparable results for all maps for the full sky and 1\% mask except for \Commander{} and \SEVEM{}, which represent outliers closer to the mean of the simulations. Excluding these two maps, we find consistent moderately significant PTE, $p = 0.04\% - 0.13\%$ ($3.2 \sigma - 3.5 \sigma$). The good agreement between the full-sky and 1\%-mask of the foreground-cleaned maps indicates that the quadrupole-octopole alignment is not significantly affected by foreground contamination. Although the $a_{\ell m}$ computed with the common mask applied are not straightforward to interpret due to the large masked fraction (26\%), we show the common-mask results for completeness. For the common mask, the $S_\mathrm{QO}$-statistic shows better agreement with the Gaussian statistically isotropic $\Lambda$CDM expectation with $p = 7\% - 14\%$ ($1.5\sigma - 1.8\sigma$).

For the $S_\mathrm{QO}$-statistic, we find negligible shifts when comparing CMB-only and cleaned-CMB simulations, which include the impact of CMB-foreground chance correlation. The PTE shifts by $\Delta p \leq 0.04\%$ for the full sky and 1\% mask and $\Delta p \leq 0.17\%$. Thus we use the $10^5$ CMB-only simulations to quote significances of $S_\mathrm{QO}$.

Our results for the Planck component-separated maps are in good agreement with \cite{Jones:2023ncn}, albeit we find slightly more significant PTE than \cite{Planck:2013lks, Muir:2018hjv} possibly due to the different maps that were used in these works. For higher-$\ell$ alignments see also \cite{Pinkwart:2018nkc, Oliveira:2018sef, Patel:2024oyj, Rodrigues:2024pkh}.

Thus, we find moderately significant PTE values in good agreement for all maps (excluding outliers \Commander{} and \SEVEM) using the full sky or 1\% mask with $p = 0.04 \% - 0.13 \%$  ($3.2\sigma - 3.5 \sigma$). The consistency between the full-sky and the 1\%-mask results suggests that foreground contamination does not significantly impact the quadrupole-octopole alignment. This confirms the moderate significance of this feature found in previous studies with improved foreground cleaning.
\vspace{7mm}


\subsection{Northern variance, $\sigma_{16}^2$}
\label{sec:sigma16}

It was pointed out that the northern (ecliptic or galactic) hemisphere at low resolution exhibits lower variance than expected from Gaussian statistically isotropic simulations, while the southern hemisphere's variance is well within expectations \citep{Eriksen:2003db, Hansen:2004mj}. 
Comparing the variance on different parts of the sky is a probe of statistical isotropy and thus the cosmological principle. The choice to probe hemispherical variance asymmetry is a posteriori.
We probe this ``low northern variance'' at $N_\side = 16$, adopting the same simple estimator as in \cite{Jones:2023ncn}:
\begin{equation}
    \sigma_{16}^2 = \overline{(T-\bar T)^2},
\end{equation}
where the bar denotes an average over all $N_\side=16$ pixels in the northern \textit{ecliptic} hemisphere. We checked that we find qualitatively similar results for the northern \textit{galactic} hemisphere. Moreover, defining $\bar{T}$ as the average over both hemispheres (with the respective mask applied) does also not alter the results.

\begin{figure}[t]
    \centering
    \includegraphics[width=0.9\linewidth]{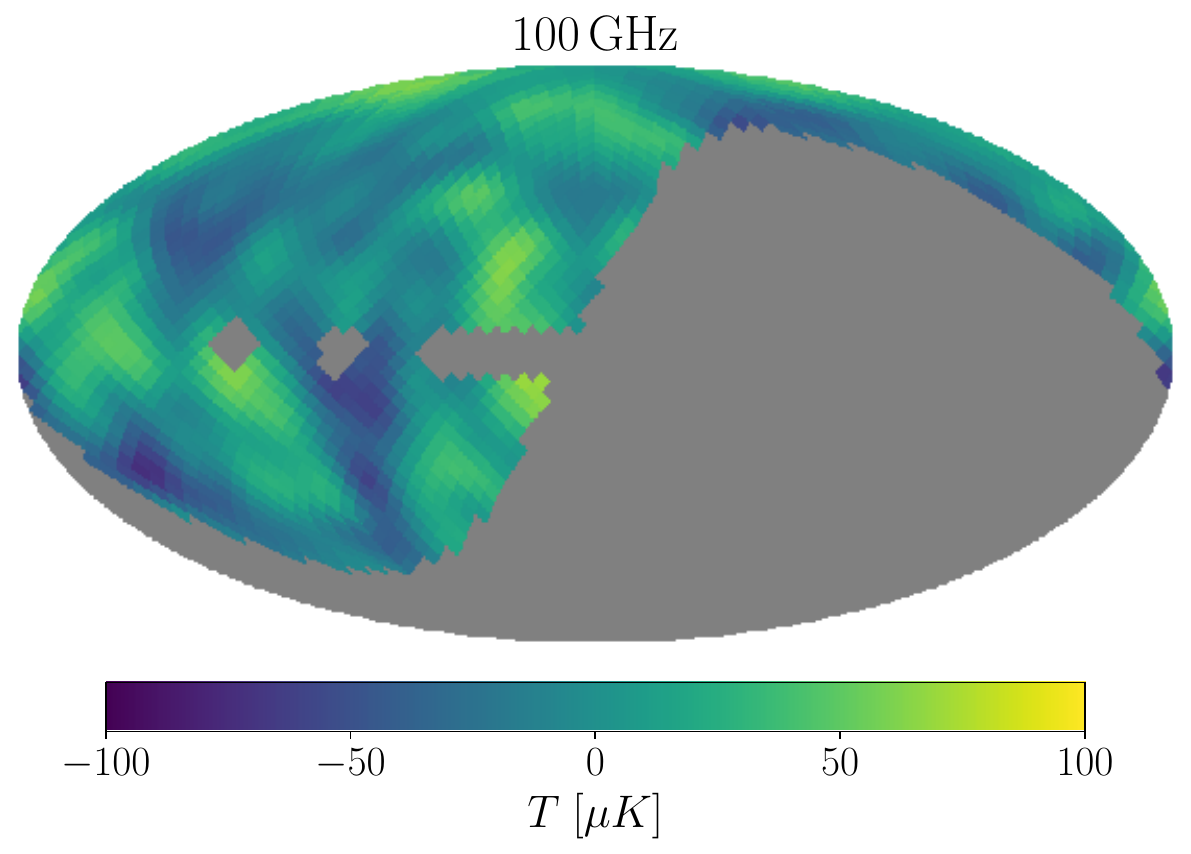}\vspace{3mm}
    \includegraphics[width=0.8\linewidth]{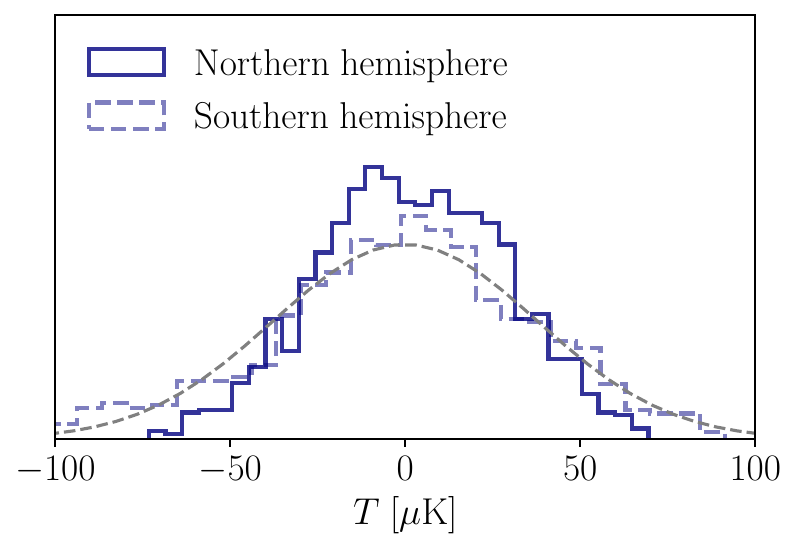}
    \caption{\textit{Top:} Foreground-cleaned $100\,\mathrm{GHz}$ map at $N_\side = 16$ with the downgraded 1\% mask (corresponding to a 3.6\% mask) and a mask covering the southern ecliptic hemisphere applied. $\sigma_{16}^2$ is defined as the variance over the unmasked pixels. \textit{Bottom:} Histogram of pixel values. The variance on the northern hemisphere (blue solid) appears smaller than the one on the southern hemisphere (blue dashed), as well as the variance expected from the Gaussian statistically isotropic simulations (gray dashed).}
    \label{fig:low_northern_variance}
\end{figure}

As an example, in the top panel of Fig.~\ref{fig:low_northern_variance}, we show the foreground-cleaned $100\,\mathrm{GHz}$ map with the 1\% mask applied at $N_\side = 16$. The 1\% mask (Planck common mask) is downgraded to $N_\side = 16$ as described in Sec.~\ref{sec:methods}, which enlarges the mask from $1.0\%$ ($25.6\%$) to $3.6\%$ ($45.5\%$). Moreover, we apply a mask that covers the southern ecliptic hemisphere. $\sigma_{16}^2$ is then simply defined as the variance of the unmasked $N_\side = 16$ pixels. The bottom panel of Fig.~\ref{fig:low_northern_variance} shows the pixel-value histograms of the same map. The variance on the northern hemisphere appears smaller than the one on the southern hemisphere. For comparison, we show a Gaussian distribution with mean and variance obtained from the Gaussian statistically isotropic $\Lambda$CDM simulations (gray dashed).

The fourth row in Fig.~\ref{fig:anomaly_stats_all} shows $\sigma_{16}^2$ for the eight maps and three sky cuts. For a given sky cut, we find that the $p$-values for all cleaning procedures are in qualitative agreement, only the \SEVEM{} full-sky result represents an outlier closer toward the mean of the simulations. 
We find that the $\sigma^2_{16}$-statistic for the common-mask case shows the lowest PTE of $p = 0.14\% - 0.17\%$ ($3.1\sigma - 3.2\sigma$) for the \cite{Nofi:2025a} foreground-cleaned maps and $p = 0.13\% - 0.36\%$ ($2.9\sigma - 3.2\sigma$) for the \cite{Planck:2018yye} component-separated maps. Using the downgraded 1\% mask leads to slightly lower significance with $p = 0.34\% - 0.40\%$ ($2.9\sigma$) for the foreground-cleaned maps and $p = 0.19\% - 0.40\%$ ($2.9\sigma-3.1\sigma$) for the component separated maps.
The full-sky results yield $p = 0.48\% - 0.68\%$ ($2.7\sigma - 2.8\sigma$) for all maps excluding the outlier \SEVEM{}. Thus we find a slight reduction of the significance when going over to a smaller galactic mask. This can be partially explained by the relatively high-temperature excursion (yellow) below the galactic center in the top panel of Fig.~\ref{fig:low_northern_variance}. Covering this region by the common mask leads to a lower variance on the northern ecliptic hemisphere. 

For the $\sigma^2_{16}$-statistic, we find negligible shifts when comparing CMB-only and cleaned-CMB simulations, which include the impact of CMB-foreground chance correlation. The PTE shifts by $\Delta p \leq 0.17\%$ for the full sky, $\Delta p \leq 0.05\%$ for the 1\% mask and by $\Delta p \leq 0.03\%$ for the common mask. Thus we use the $10^5$ CMB-only simulations to quote significances of $\sigma^2_\mathrm{16}$. 

Our results for the low northern variance under the \cite{Planck:2018yye} component-separated maps agree well with the literature \citep{Planck:2019evm, Jones:2023ncn}. 

In summary, we find moderately significant PTE values in qualitative agreement for all maps and sky cuts. We find a slight reduction in significance when going over from the common mask ($3.1\sigma - 3.2\sigma$) to the smaller 1\% mask ($2.9\sigma$) for the foreground-cleaned maps. Note, however, that the selection of the northern hemisphere is an arbitrary a-posteriori choice and that the northern (galactic or ecliptic) hemisphere does not have a meaning for the CMB-only simulations. Regardless, we are imposing the same hemispherical coordinate cut to the simulations. In principle, this motivates a look-elsewhere correction, which would reduce the significance of this statistic.


\subsection{Local-variance asymmetry, $A_\mathrm{LV}$}
\label{sec:ALV}

While $\sigma^2_{16}$ measures the variance of the entire northern hemisphere at low resolution (excluding masked areas), another common measure of hemispherical asymmetry considers the \emph{local} variance in small disks. This is motivated by the observation that there seems to be a general asymmetry in power between the two hemispheres in WMAP and Planck data \citep{Eriksen:2003db, Hansen:2004mj, Planck:2013lks, Akrami:2014eta}.
Here, we compute the local variance in disks on the sky and measure the amplitude of a dipole in such a ``local-variance map'' following the approach in previous works. The (a)symmetry of such a local-variance map is a probe of the statistical isotropy and thus the cosmological principle. Compared to the $\sigma_{16}$-statistic, this statistic fits for the direction that maximizes local-variance asymmetry.

The local variance in a disk of radius $\theta$ centered on the pixel position $\hat n$ can be defined as \citep{Akrami:2014eta, Planck:2015igc}:
\begin{equation}
    \label{eq:ALV}
    \sigma_\theta^2(\hat{n}) = \frac{1}{N_{\mathcal{D}_{\theta}(\hat{n})}} \sum_{i\in \mathcal{D}_\theta(\hat{n})} \big[ T(\hat n_i) - \bar{T}_\theta(\hat n)\big]^2,
\end{equation}
where $\mathcal{D}_{\theta}(\hat{n})$ is the set of unmasked pixels within the disk and $N_{\mathcal{D}_{\theta}(\hat{n})}$ their number. Note that this measures the variance with respect to the mean temperature, $\bar{T}_\theta(\hat n)$, of the unmasked pixels \textit{within the disk}, and not with respect to the mean CMB temperature. The statistic in Eq.~\eqref{eq:ALV} is thus not a generalization of $\sigma^2_{16}$ (Sec.~\ref{sec:sigma16}) but an entirely different statistic. We found that instead defining the variance with respect to the mean full-sky or cut-sky CMB temperature leads to different conclusions; such a statistic computed from the CMB maps considered in this works is in good agreement with the expectations of a Gaussian statistically isotropic $\Lambda$CDM sky (with PTE of the order of tens of percent).

\begin{figure}
    \centering
    \includegraphics[width=1.\linewidth]{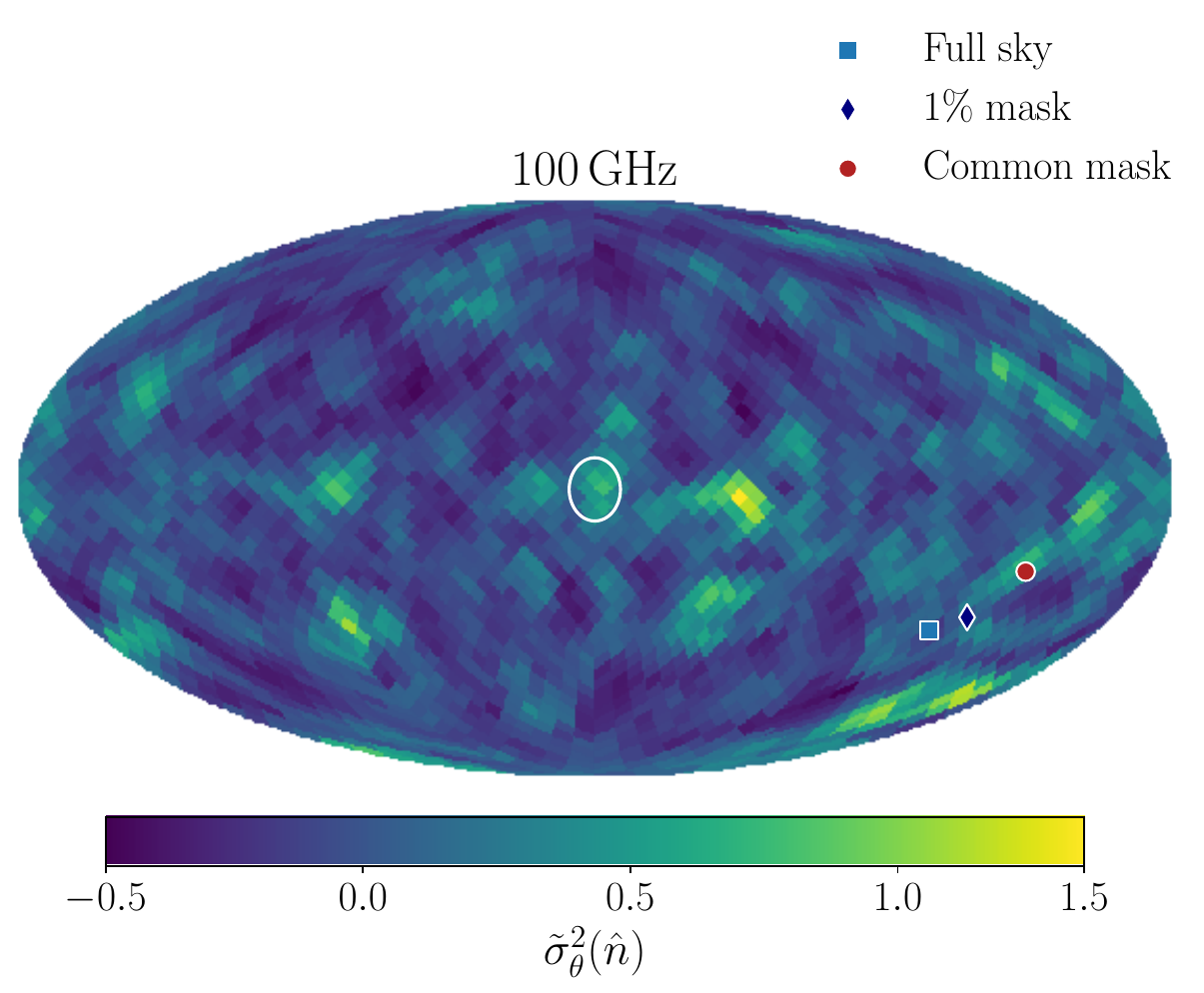}
    \caption{Normalized local-variance map, $\tilde{\sigma}_\theta^2(\hat{n})$ (Eq.~\eqref{eq:ALV_normalized}) of the $100\ \mathrm{GHz}$ measured within $8^\circ$ disks (illustrated as the white circle in the center). The significance of the hemispherical asymmetry is quantified via the amplitude, $A_\mathrm{LV}$, of the dipole of this map, which is around $2\sigma$ for the full sky and 1\% mask and around $3\sigma$ for the Planck common mask. The local-variance dipole directions for the $100\, \mathrm{GHz}$ map are indicated as the markers for the three different sky cuts as indicated in the legend.}
    \label{fig:norm_LV_map_100GHz}
\end{figure}

Following \cite{Muir:2018hjv}, we normalize the local-variance map via
\begin{equation}
    \label{eq:ALV_normalized}
    \tilde{\sigma}_\theta^2(\hat{n}) = \frac{\overline{w}}{w(\hat n)}\, \frac{\sigma_\theta^2(\hat{n}) - \mu_{\sigma_\theta}(\hat n)}{\mu_{\sigma_\theta}(\hat n)},
\end{equation}
where $\mu_{\sigma_\theta}(\hat n)$ is the mean of the local-variance maps obtained from the Gaussian statistically isotropic CMB simulations. The dimensionless weight map is defined as
\begin{equation}
    w(\hat n) = \frac{1}{N_\mathrm{sim}} \sum_{\alpha = 1}^{N_\mathrm{sim}} \left[\frac{\sigma_\theta^{2(\alpha)}(\hat n) - \mu_{\sigma_\theta}(\hat n)}{\mu_{\sigma_\theta}(\hat n)}\right]^2,
\end{equation}
where $\alpha$ labels the $N_\mathrm{sim} = 10^5$ CMB simulations. The weight map $w \sim \mathrm{Var}[\sigma_\theta^2(\hat{n})]$ is thus a measure of the variance of the simulated local-variance maps. We compute local-variance maps from CMB maps at $N_\mathrm{side} = 128$ by centering disks of radius $\theta=8^\circ$ at the pixels of an $N_\mathrm{side} = 16$ map. $\theta=8^\circ$ represents a common choice in the literature since it gives comparatively significant results \citep{Planck:2013lks, Muir:2018hjv}. Following previous works, we construct a mask of the normalized local-variance map by defining an output pixel as masked if more than $90\%$ of pixels are masked. This results in $0\%$ ($11\%$) masked pixels for the 1\% mask (common mask). As an example, we show the normalized local-variance map obtained from the cleaned $100\ \mathrm{GHz}$ map with the 1\% mask applied in Fig.~\ref{fig:norm_LV_map_100GHz}.

The amplitude $A_\mathrm{LV}$ of a possible dipole in the local-variance map defined in Eq.~\eqref{eq:ALV_normalized} is then obtained via \texttt{healpy's remove\_dipole}. The preferred directions of this local-variance dipole are shown as the markers in Fig.~\ref{fig:norm_LV_map_100GHz} for the $100\ \mathrm{GHz}$ map; the other maps are shown in App.~\ref{app:all}. With the Planck common mask applied, the directions of the local-variance dipoles of all maps agree well. For the other sky cuts, we find good agreement of the local-variance-dipole directions with the common-mask case, except for \Commander{} and \SEVEM{} for the full-sky and $1\%$-mask and except for the $94\,\mathrm{GHz}$ map for the 1\%-mask case. For these outliers, the local-variance dipoles point toward the galactic plane, which can be explained by galactic foreground contamination that leads to enhanced local variance. We exclude these maps for the respective sky cuts in the discussion below. 

We present the results for the local-variance amplitude, $A_\mathrm{LV}$, in the bottom row of Fig.~\ref{fig:anomaly_stats_all}. For the common mask, we find good agreement for all maps and we obtain moderately significant PTE values of $p = 0.13\% - 0.16\%$ ($3.2\sigma$) for the foreground-cleaned maps and $p = 0.10\% - 0.11\%$ ($3.3\sigma$) for the component-separated maps. For the 1\% mask, we find less significant deviations from the mean of the simulations with a larger spread between the cleaning procedures with $p = 2.2\% - 2.8\%$ ($2.2\sigma-2.3\sigma$) for the foreground-cleaned maps and $p = 0.3\% - 0.8\%$ ($2.7\sigma - 3.0\sigma$) for the component-separated maps, excluding \Commander{} and \SEVEM{}. For the full sky, we find slightly less significant PTE than for the other two sky cuts, with $p = 4.0\% - 8.6\%$ ($1.7\sigma - 2.1\sigma$, excluding the $94\, \mathrm{GHz}$ map) and $p = 0.8 \% - 0.9\%$ ($2.6\sigma - 2.7\sigma$, excluding \Commander{} and \SEVEM). The difference between the foreground-cleaned maps and the two component-separated maps could be caused by inpainting techniques in the component-separated maps (\SMICA, \NILC), which could reduce the local variance in the galactic-plane region. 

We compare the PTE obtained with the CMB-only and cleaned-CMB simulations, in order to probe the impact of CMB-foreground chance correlation. The PTE shifts by $\Delta p \leq 4.3\%$ for the full sky,  $\Delta p \leq 2.5\%$ for the 1\% mask and $\Delta p \leq 0.01\%$ for the common mask. The shift in PTE for the full-sky and 1\%-mask case is not small but since the $A_\mathrm{LV}$-statistic is not in the tail and $p \sim \mathcal{O}(\%)$, we use the $10^5$ CMB-only simulations to quote significances of $A_\mathrm{LV}$ for both of these cases.

Our PTE values agree well with the ones in \citep{Akrami:2014eta, Planck:2015igc, Sanyal:2024iyv}, however, we find more significant PTE than \cite{Muir:2018hjv}, which might be due to different map versions.

In summary, we find that the $A_\mathrm{LV}$-statistic for the \cite{Nofi:2025a} foreground-cleaned maps takes on moderately significant values only for the common mask ($3.2\sigma$) confirming previous results, while these are only mildly significant for the 1\% mask ($2.2\sigma - 2.3 \sigma$) and full sky ($1.7\sigma - 2.1\sigma$).


\section{Conclusions}
\label{sec:conclusions}

In this work, we used the maps obtained in \cite{Nofi:2025a} with a cleaning methodology tailored to remove foreground contamination from low-resolution Planck and WMAP maps in order to re-assess the large-scale CMB anomalies. This gave us several advantages compared to previous studies of the CMB anomalies: the improved cleaning at low resolution allows us to use more sky and thus gives better statistics; the smaller mask maintains approximate orthogonality of the $a_{\ell m}$ of different $\ell$, which is not given for the larger Planck common mask; being able to use a smaller galactic mask let us explore the dependence on the size of the galactic mask; and lastly the four maps based on an alternative cleaning procedure than the Planck component-separated maps \Commander, \NILC, \SEVEM, \SMICA, allowed us to explore the impact of different foreground-removal methods. With significances quoted for the four \cite{Nofi:2025a} foreground-cleaned maps, our main results for the five anomaly tests considered in this work are:
\begin{enumerate}
    \item Low correlation, $S_{1/2}$: We find a moderate impact of the cleaning with the foreground-cleaned maps being slightly more consistent with the Gaussian statistically isotropic $\Lambda$CDM expectation. More importantly, we find a significant impact of the choice of masking. Applying the common mask, which covers $26\%$ of the sky and is used in previous works, leads to moderately significant PTE ($p = 0.19\% - 0.24 \%$, $3.0\sigma - 3.1 \sigma$). However, applying the $1\%$ mask, enabled by the improved foreground cleaning, reduces the significance of $S_{1/2}$ ($p= 5.8\% - 7.3\%$, $1.8 \sigma - 1.9\sigma$).
    \item Parity asymmetry, $R^{27}$: We find only a mild impact of the cleaning and masking choices for this statistic (excluding the outlier maps \Commander{} and \SEVEM). We find only mildly significant PTE ($p= 2.8\% - 5.9 \%$, $1.9\sigma - 2.2 \sigma$) for all sky cuts, consistent with previous literature. Moreover, the specific choice of $\ell_\max = 27$ is chosen to maximize the significance; a correction for this look-elsewhere effect would further decrease the significance of this statistic.
    \item Multipole alignments, $S_\mathrm{QO}$: We find only a mild impact of the cleaning procedure (excluding the outlier maps \Commander{} and \SEVEM). We find good agreement between the full sky and 1\% mask with moderately significant PTE ($p = 0.05\% - 0.13 \%$, $3.2\sigma - 3.5\sigma$), confirming previous works based on full-sky maps. 
    \item Northern variance, $\sigma^2_{16}$: We find only a mild impact of the cleaning (excluding the outlier \SEVEM) but some impact of the masking: the results applying the 1\% mask are slightly more consistent with the Gaussian statistically isotropic $\Lambda$CDM expectation ($p = 0.34 \% - 0.40\%$, $2.9\sigma$) than the results applying the common mask ($p = 0.14\% - 0.17\%$, $3.1\sigma - 3.2\sigma$).
    \item Local-variance asymmetry, $A_\mathrm{LV}$: We find a moderate impact of the cleaning with the foreground-cleaned maps being slightly more consistent with the Gaussian statistically isotropic $\Lambda$CDM expectation (excluding outliers \Commander, \SEVEM, and $94\, \mathrm{GHz}$). However, we find a notable impact of the masking: the 1\%-mask results are more consistent with the $\Lambda$CDM expectation ($p = 2.2\% - 2.8\%$, $2.2\sigma - 2.3 \sigma$) than the common-mask results ($p = 0.13\% -0.16\%$, $3.2\sigma$). 
\end{enumerate}
Overall, we find consistent results for most anomaly statistics for the different foreground-cleaning procedures with some impact of the cleaning procedure on $S_{1/2}$ and $A_\mathrm{LV}$. However, the choice of foreground mask shows a notable impact on $S_{1/2}$, $A_\mathrm{LV}$, and a mild impact on $\sigma^2_{16}$, with the smaller 1\% mask leading to a lower significance of the statistics then the 26\% Planck common mask. 
For $S_\mathrm{QO}$ and $R^{27}$ we confirm previous results while including a larger sky fraction. Our results based on maps with improved foreground cleaning indicate that the CMB anomalies are not caused by galactic foregrounds, however, some are affected by the choice of galactic mask.

The anomaly statistics considered in this work were defined a posteriori, i.e.\ only after the features were observed in the data. In principle, this warrants a correction for the look-elsewhere effect, which accounts for the fact that the probability of finding a seemingly significant result for one specific test increases with the number of possible tests. However, in practice it is difficult to implement a complete look-elsewhere correction since it is not straightforward to determine the number of all possible anomaly tests for the CMB. The simplest probe of the look-elsewhere effect is to explore the dependence of significance on the exact definition of a statistic. We explored the dependence of $S_\mu$ on the integration limit $\mu$ and the dependence of the $R^{\ell_\max}$ on the maximum multipole, $\ell_\max$. For both statistics, we find a range of values that give comparable PTE than the common choices, $\mu =0.5$ and $\ell_\max = 27$, respectively, although outside of these ranges, the real maps are more consistent with the Gaussian statistically isotropic expectation. Thus in order to assess an accurate estimate of the significance, a more thorough study of the look-elsewhere effect is necessary, which would lower the significance of the individual anomaly tests. 

While we find a notably lower significance for $S_{1/2}$ and $A_\mathrm{LV}$ for the foreground-cleaned maps and mask (around $2\sigma$), we find broad qualitative agreement between our results and previous work for the remaining three statistics. For these statistics, we confirm the moderate significance of roughly $2\sigma$ to $3\sigma$.
Thus the significance of each individual statistic is not significant enough to necessarily warrant new physics. 
While we do not have an alternate model to allow for a direct hypothesis test, such a test cannot strongly prefer an alternate model given the moderate degree to which the individual anomalies are disfavored by $\Lambda$CDM, even before look-elsewhere corrections. In order for a model to be significantly preferred over $\Lambda$CDM, it would need to explain several of the anomalies simultaneously or provide a better description of some other measurement (separate from large-scale CMB temperature fluctuations). 

In this work, we explored the impact of the foreground-cleaning procedure and masking choice on commonly studied anomalies of the CMB temperature maps. Future CMB data from polarization \citep[e.g.][]{Shi:2022hxc, Banday:2025scr} or measurements of the remote quadrupole \citep[e.g.][]{Kamionkowski:1997na, Deutsch:2017ybc} might be able to provide complementary information about the CMB anomalies. 

\vspace{5mm}
Code to reproduce the results in this work is available in a Zenodo repository \citep{Herold:19699501}.

\vspace{5mm}
\textit{Acknowledgments:} LH is grateful for helpful discussions about the CMB anomalies with Adrienne Erickcek, Craig Hogan, Dragan Huterer, Joann Jones, Stephan Meyer, Rui Shi, and Glenn Starkman. LH was supported by a William H. Miller fellowship. This research was supported by NASA grant 80NSSC24K0625. We acknowledge the use of the Legacy Archive for Microwave Background Data Analysis (LAMBDA), part of the High Energy Astrophysics Science Archive Center (HEASARC). HEASARC/LAMBDA is a service of the Astrophysics Science Division at the NASA Goddard Space Flight Center. We also acknowledge the use of the Planck Legacy Archive. Planck is an ESA science mission with instruments and contributions directly funded by ESA Member States, NASA, and Canada.

\vspace{5mm}
\textit{Software:} HEALPix \citep{gorski/etal:2005, healpy:2019}, matplotlib \citep{hunter:2007}, numpy \citep{harris/etal:2020}, Polspice \citep{Szapudi:2000xj, Chon:2003gx, Challinor:2011}, polyMV \citep{Oliveira:2018sef}, scipy \citep{virtanen/etal:2020}.


\newpage
\appendix

\section{Anomaly tests for all maps}
\label{app:all}

\begin{figure}[h]
    \centering
    \includegraphics[width=\linewidth]{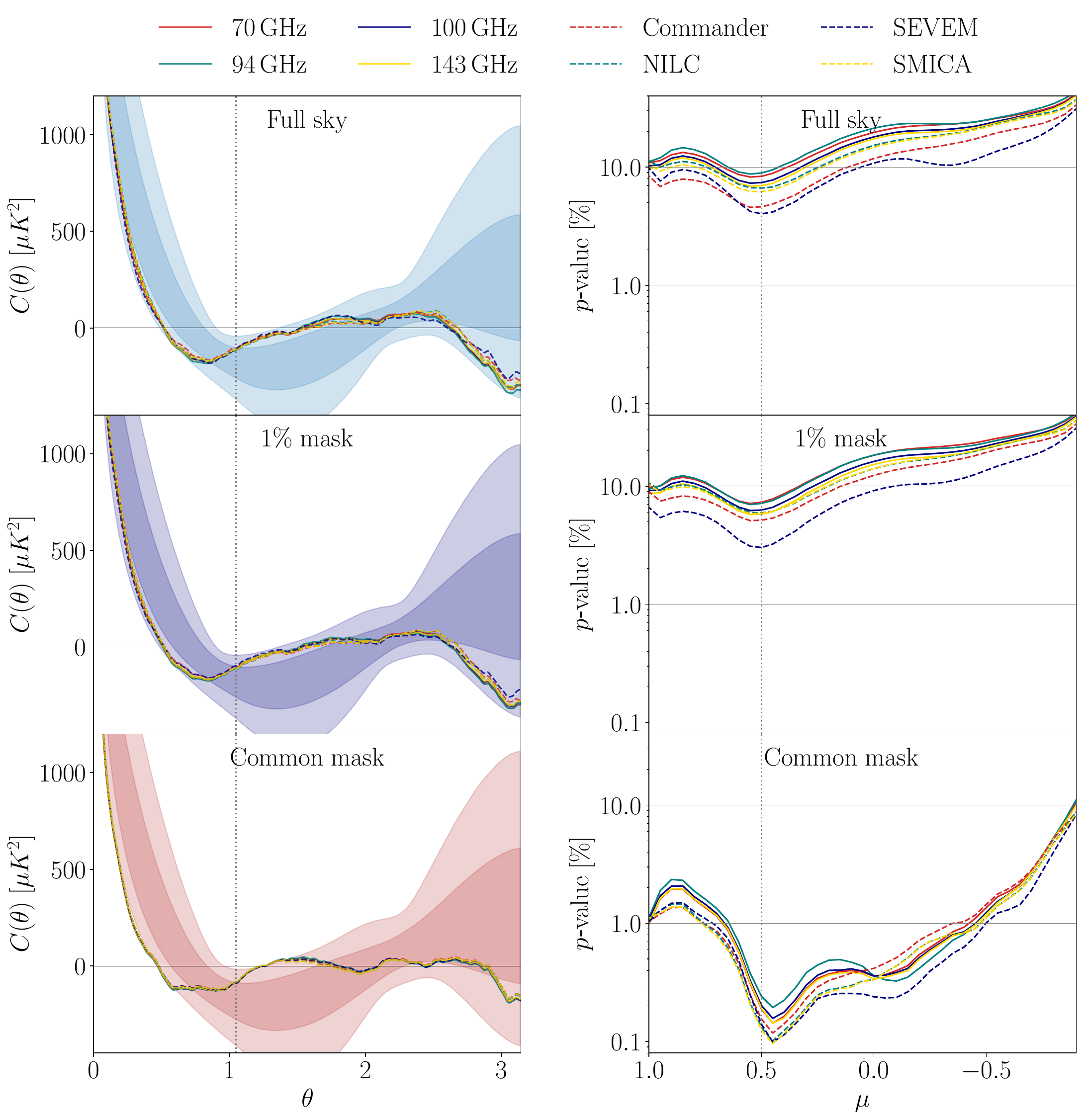}
    \caption{\textit{Left: } Angular correlation functions, $C(\theta)$, for three different sky cuts (rows) and for all eight maps (as indicated in the legend). When applying the common mask, $C(\theta)$ is closer to zero for $\theta>60^\circ$ (vertical dotted line) than expected from the $\Lambda$CDM simulations ($1\,\sigma$ and $2\,\sigma$ red shaded band); less so for the 1\% mask and full sky. \textit{Right: } Significance in $p$-values of the $S_\mu$-statistic, Eq.~\eqref{eq:S_mu}, as a function of the integration limit, $\mu = \cos(\theta)$. We find good agreement between the different cleaning procedures.}
    \label{fig:Smu_all}
\end{figure}

\begin{figure}[h]
    \centering
    \includegraphics[width=\linewidth]{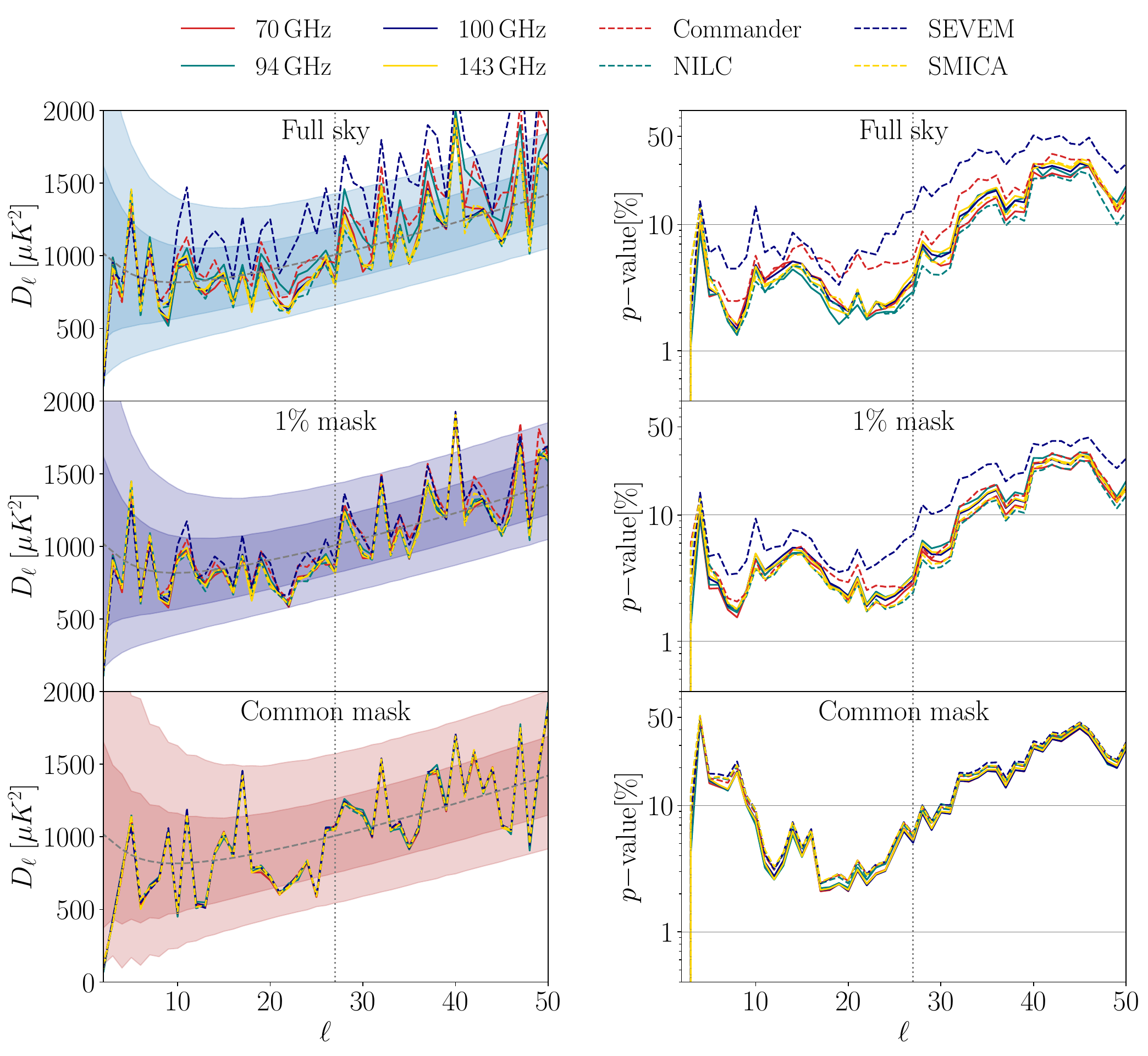}
    \caption{\textit{Left:} Power spectra, $D_\ell$, for three sky cuts (rows) and all eight maps (as indicated in the legend). \textit{Right:} $p$-values of $R^{\ell_\max}$ as a function of $\ell_\max$. The baseline choice $\ell_\max = 27$ (vertical dotted line) gives slightly more significant $p$-values than higher values of $\ell_\max$. We find good agreement between the different sky cuts and cleaning procedures (except for \Commander{} and \SEVEM{}, which represent outliers for the full sky and 1\% mask).}
    \label{fig:Rlmax_all}
\end{figure}

\begin{figure}[h]
    \centering
    \includegraphics[width=0.45\linewidth]{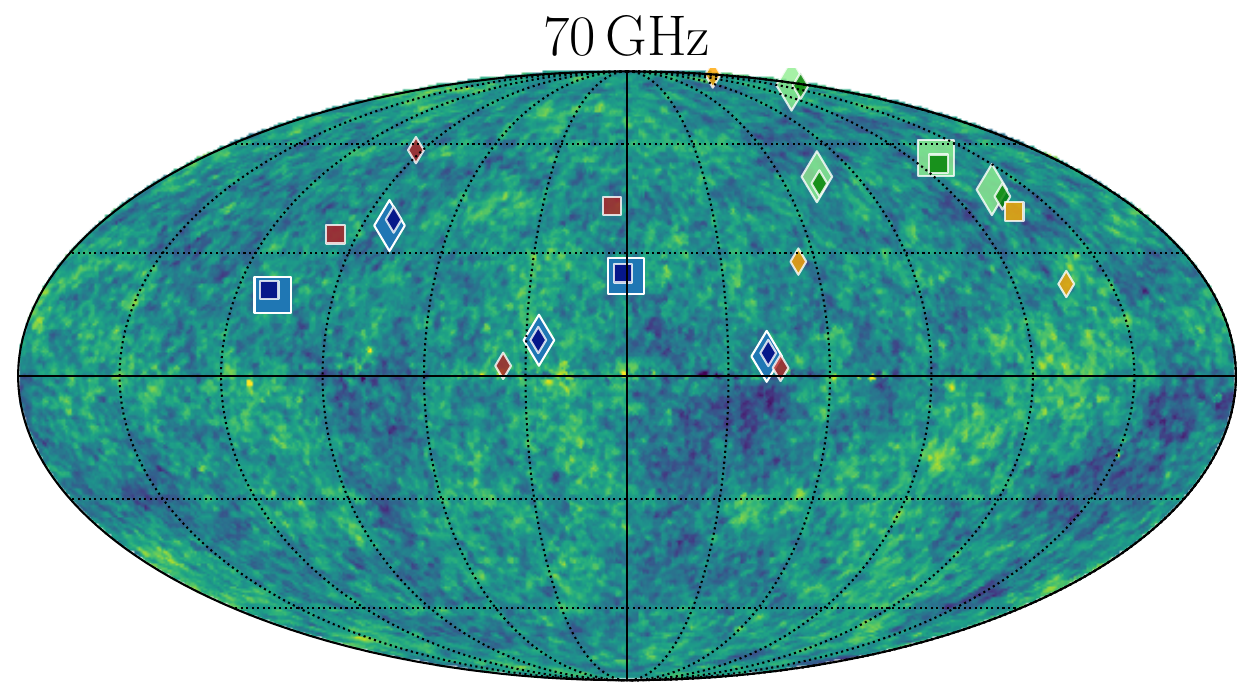}
    \includegraphics[width=0.45\linewidth]{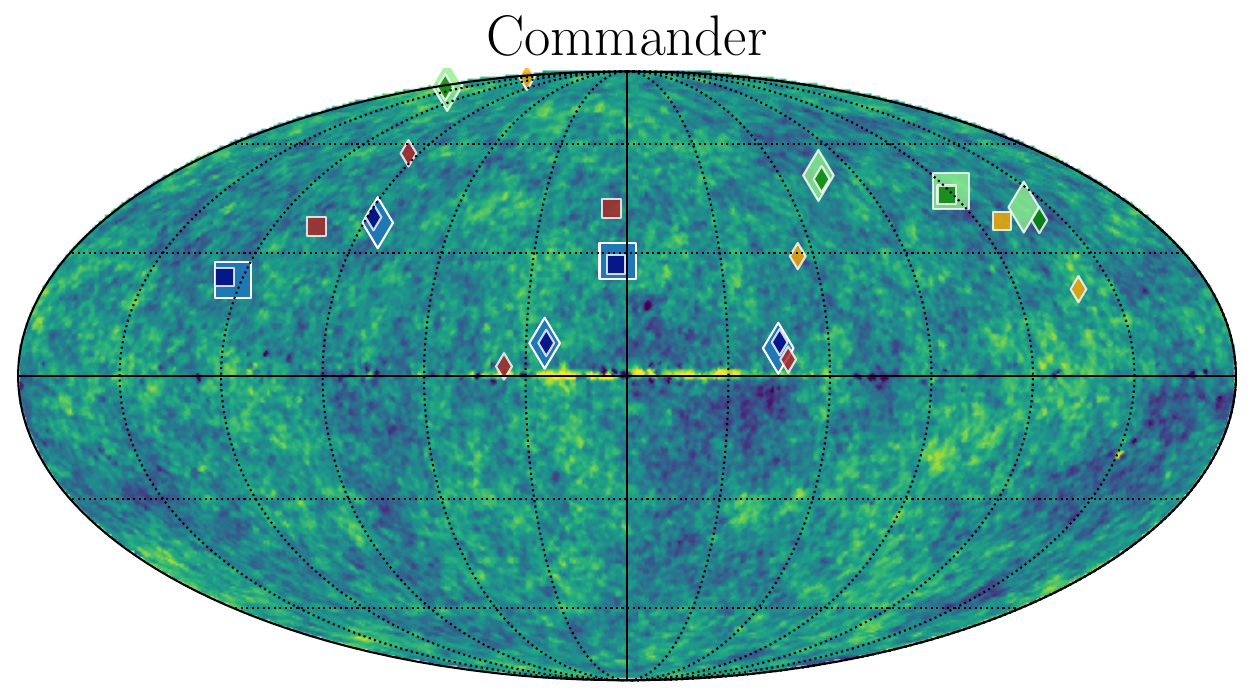}
    \includegraphics[width=0.45\linewidth]{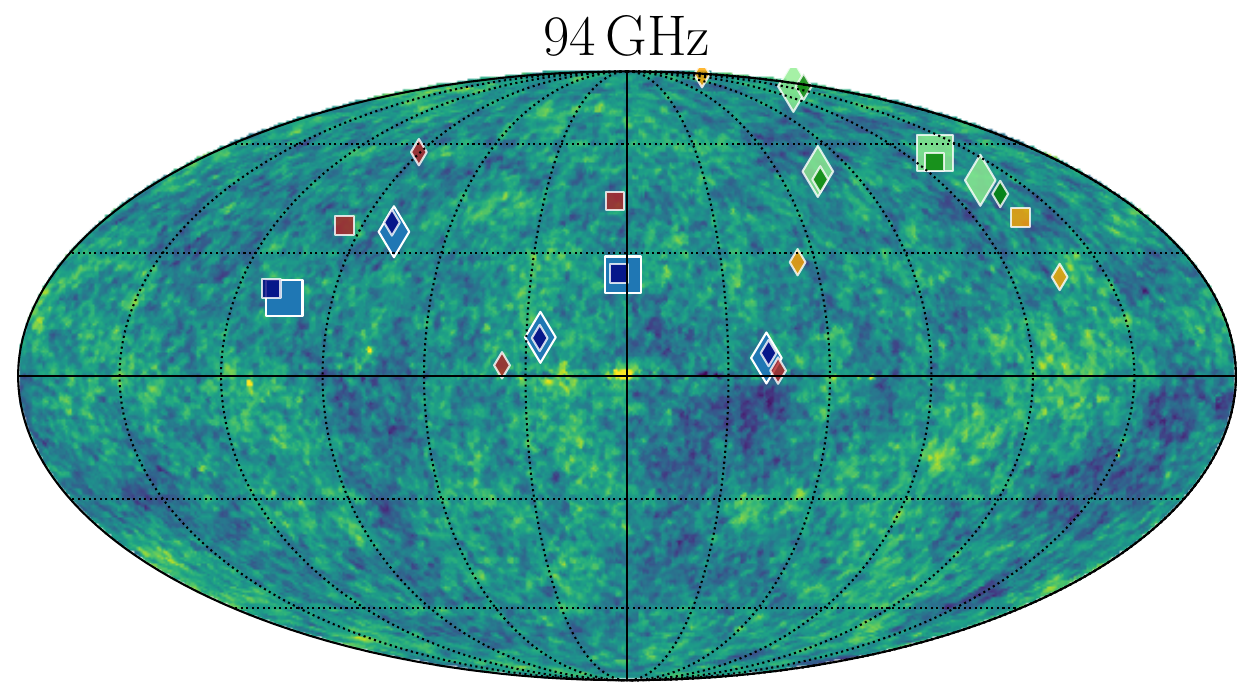}
    \includegraphics[width=0.45\linewidth]{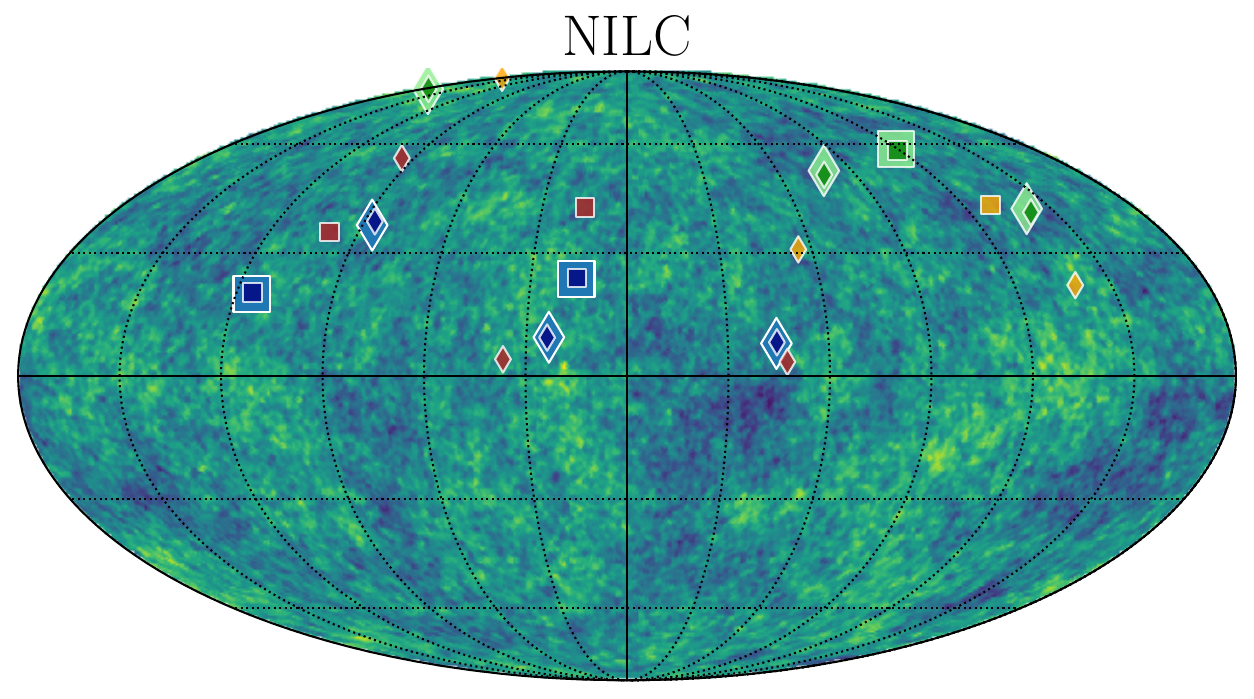}
    \includegraphics[width=0.45\linewidth]{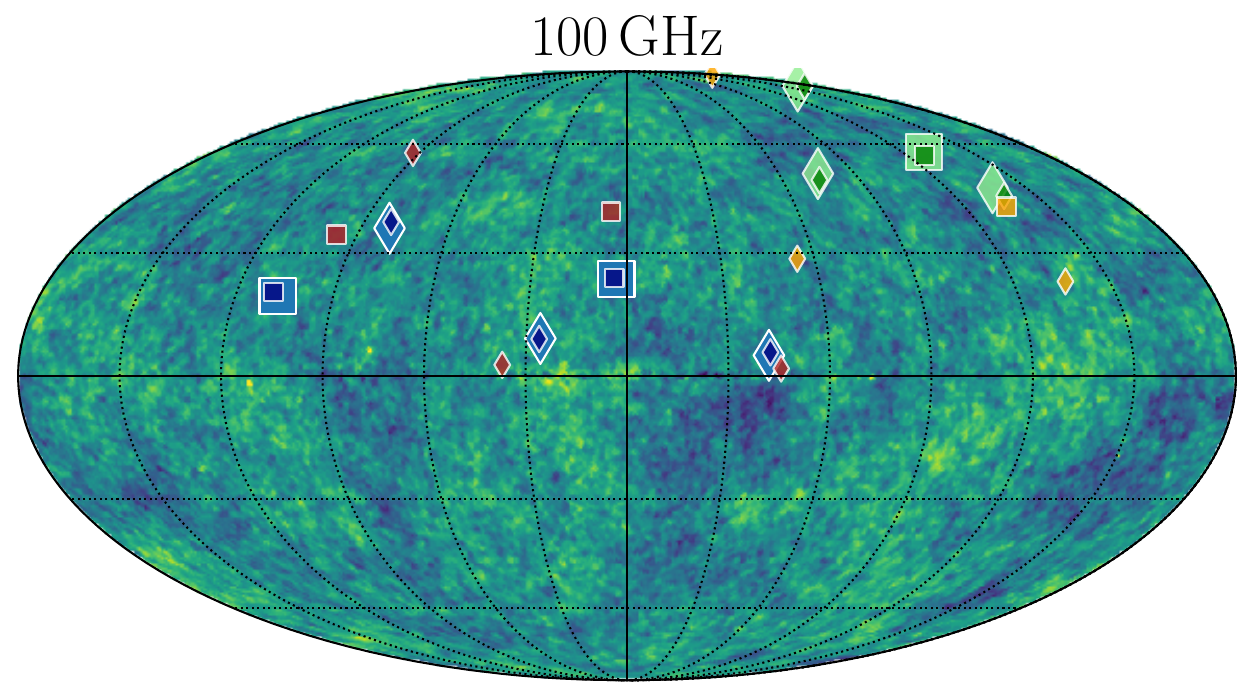}
    \includegraphics[width=0.45\linewidth]{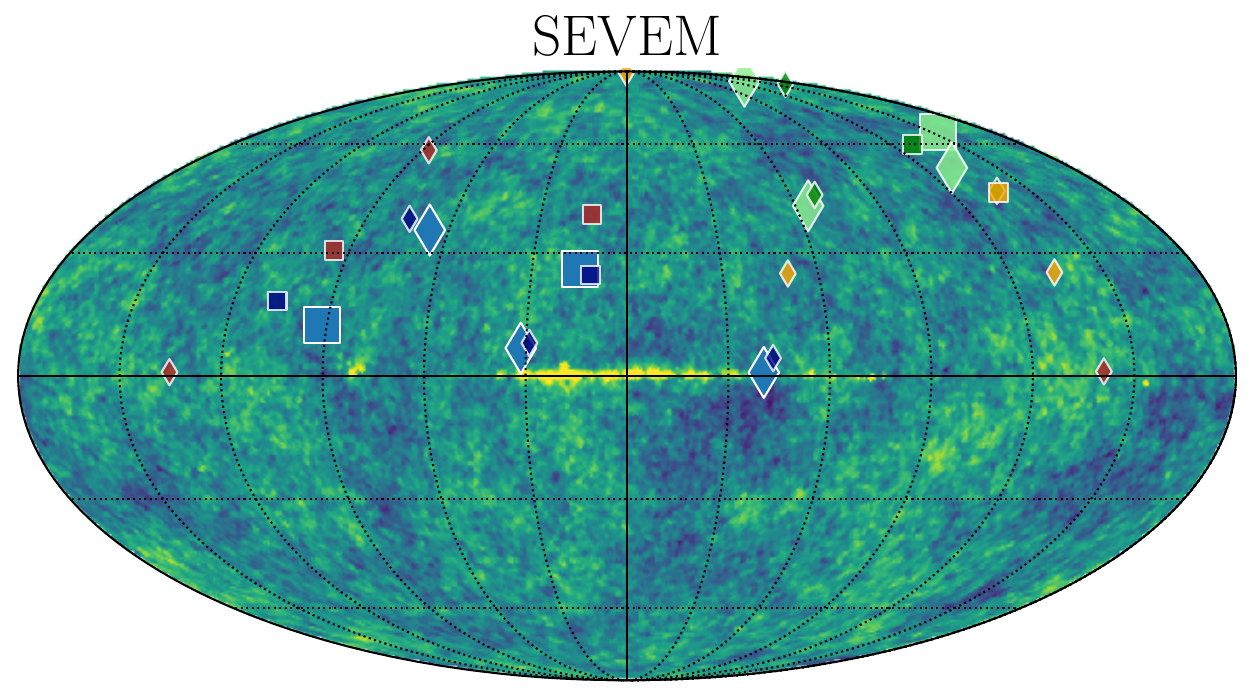}
    \includegraphics[width=0.45\linewidth]{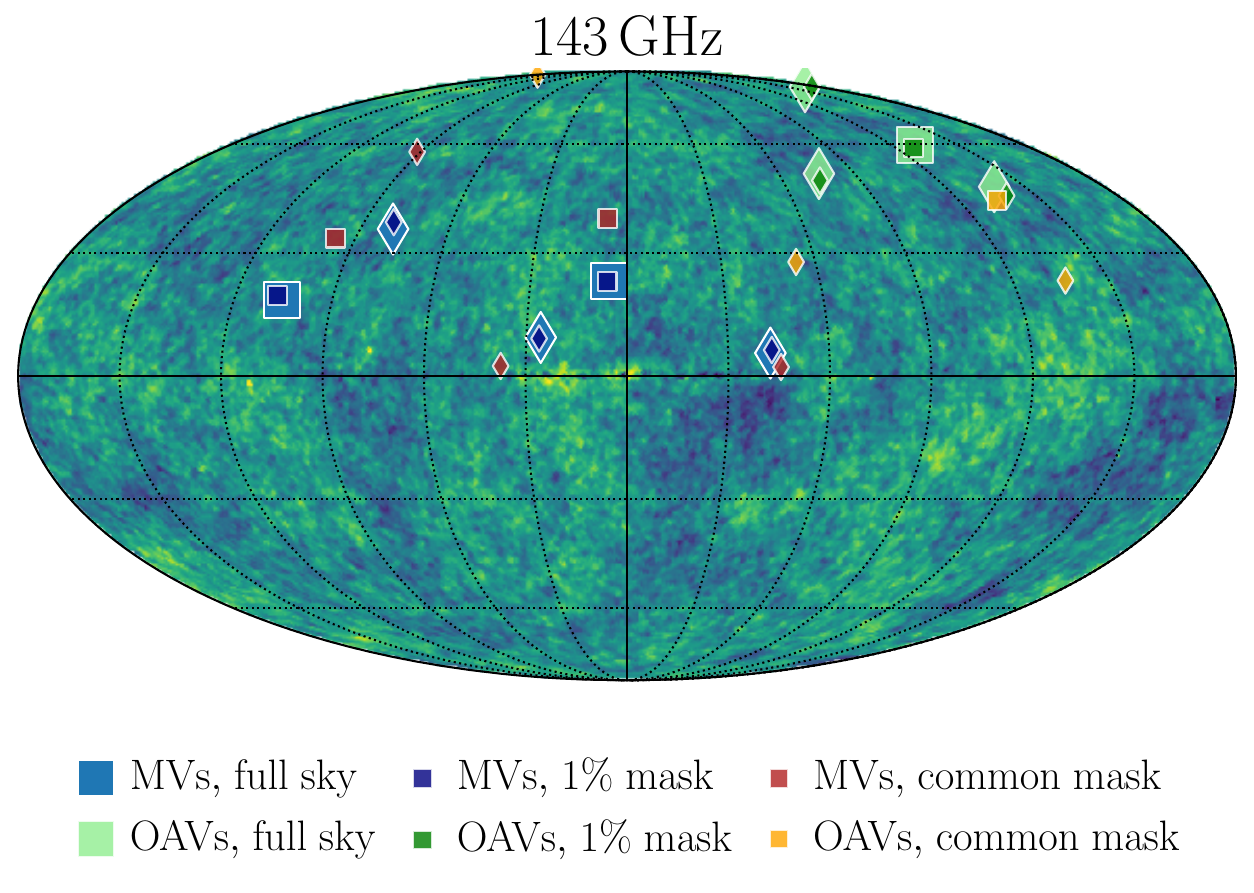}
    \includegraphics[width=0.45\linewidth]{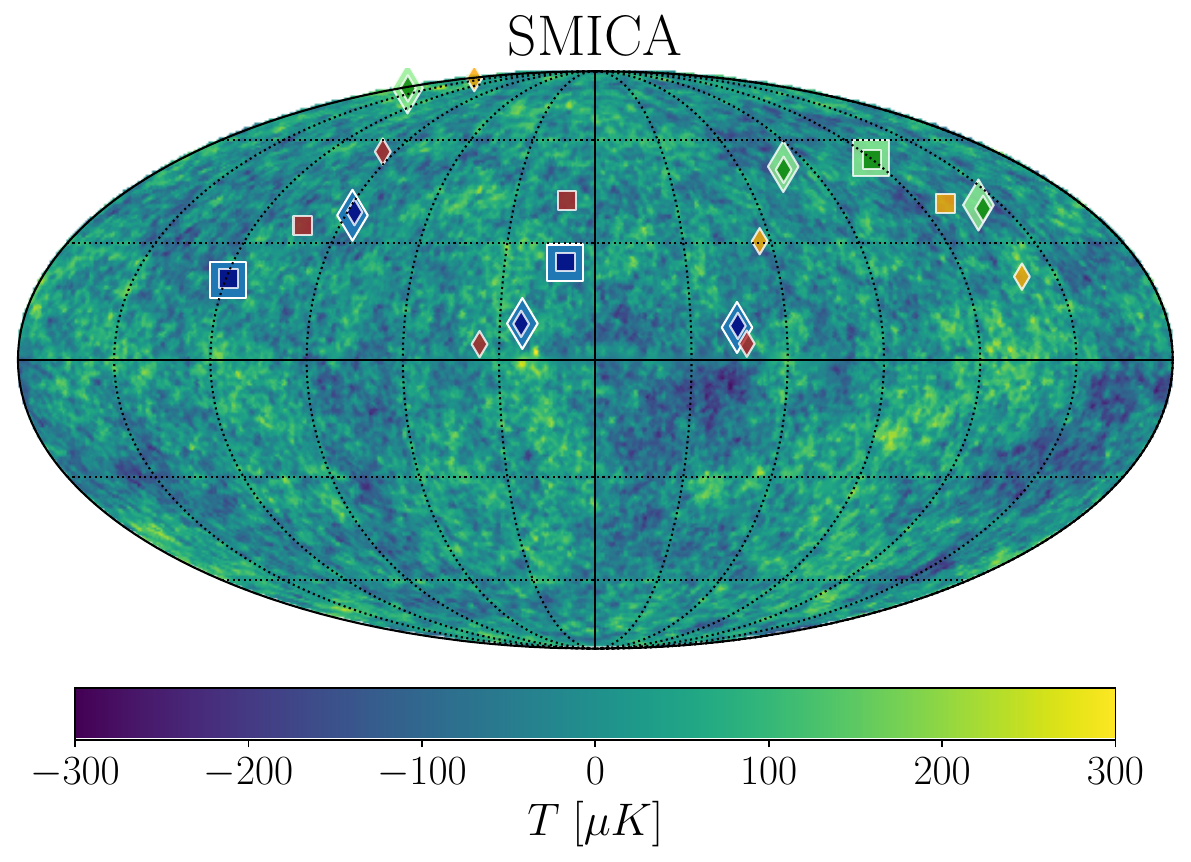}
    \caption{Multipole vectors (MVs) and oriented area vectors (OAVs) of the quadrupole (squares) and octopole (diamonds) obtained for the three sky cuts (as indicated in the legend) for all eight maps. The temperature maps at $N_\mathrm{side} = 128$ are shown in the background. For all maps and the full-sky and 1\%-mask case, we recover the approximate alignment of the planes spanned by the MVs of quadrupole and octopoles. The alignment is weakened when applying the common mask, whose MVs are less straightforward to interpret (see text).}
    \label{fig:SQO_all}
\end{figure}

\begin{figure}[h]
    \centering
    \includegraphics[width=0.45\linewidth]{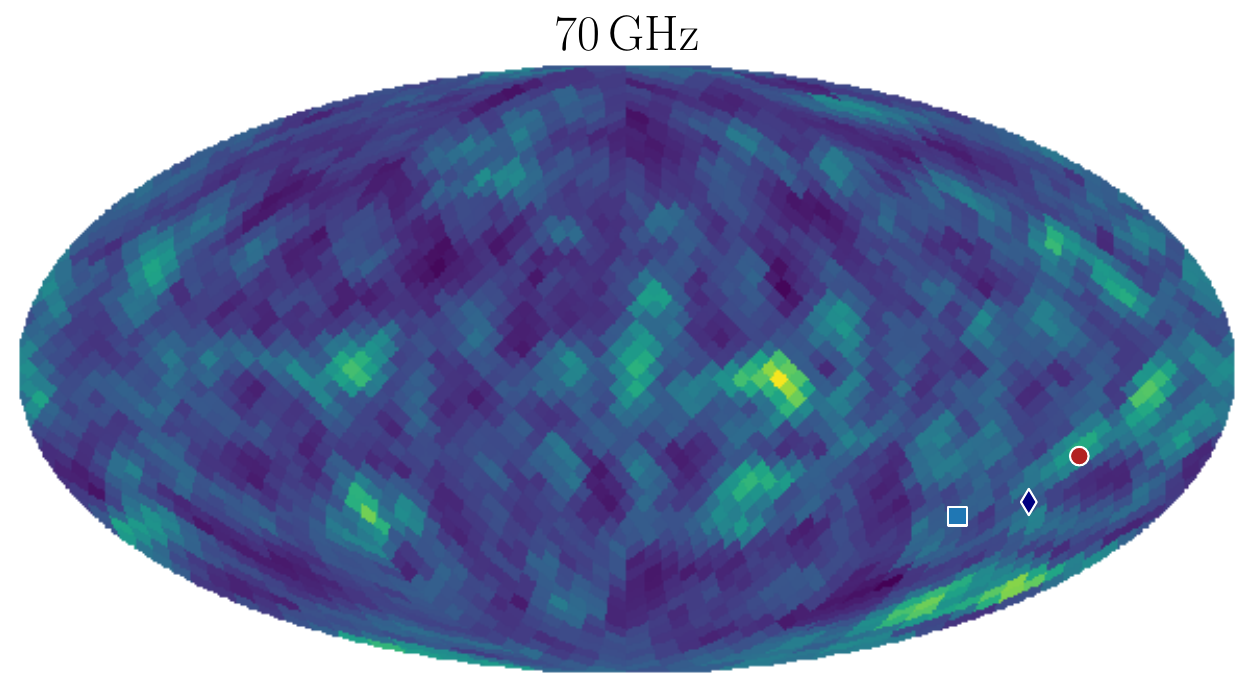}
    \includegraphics[width=0.45\linewidth]{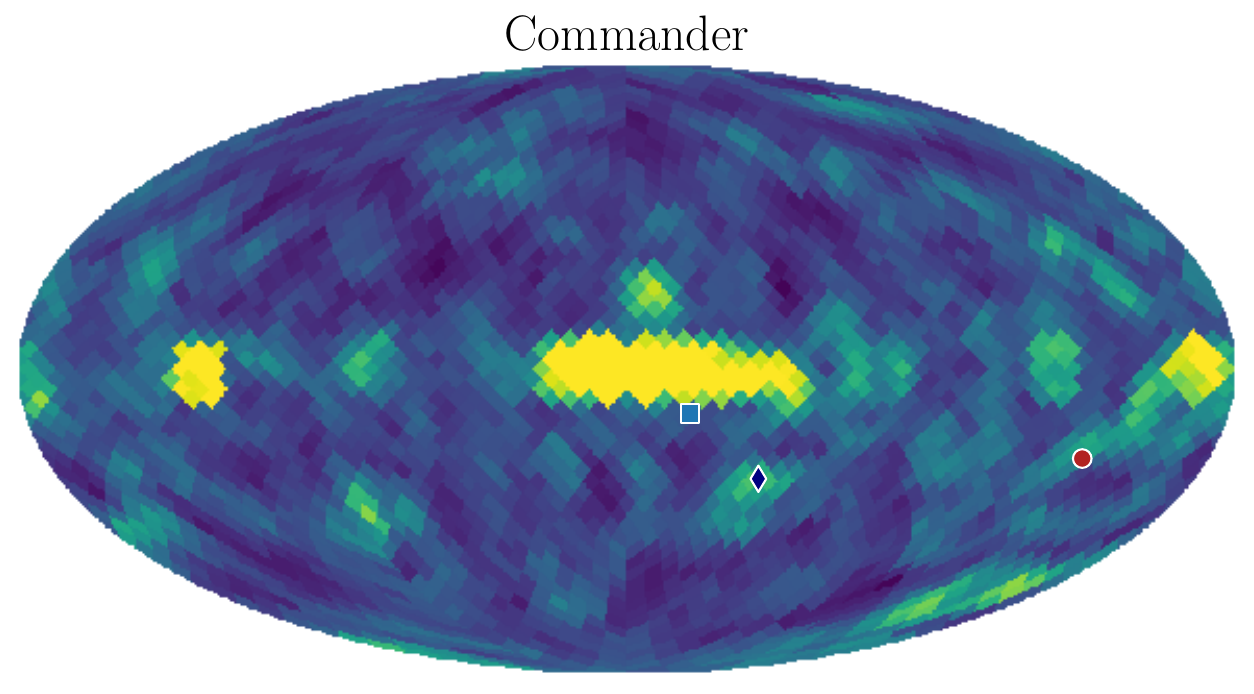}
    \includegraphics[width=0.45\linewidth]{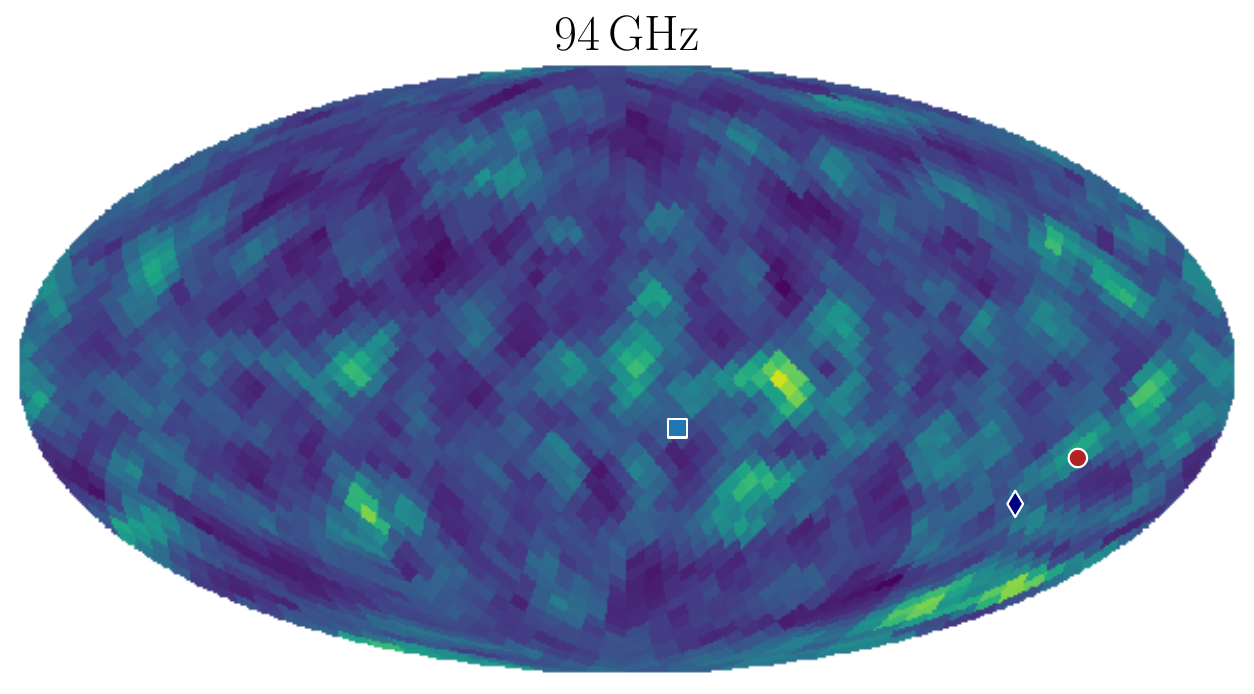}
    \includegraphics[width=0.45\linewidth]{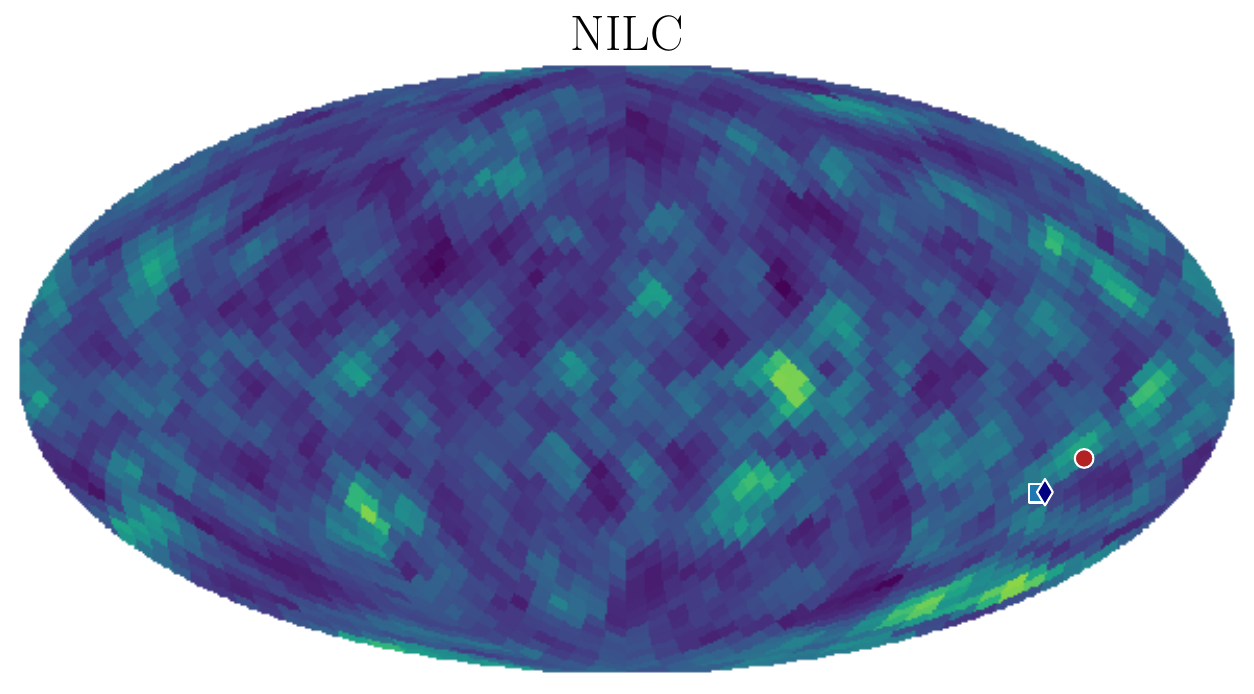}
    \includegraphics[width=0.45\linewidth]{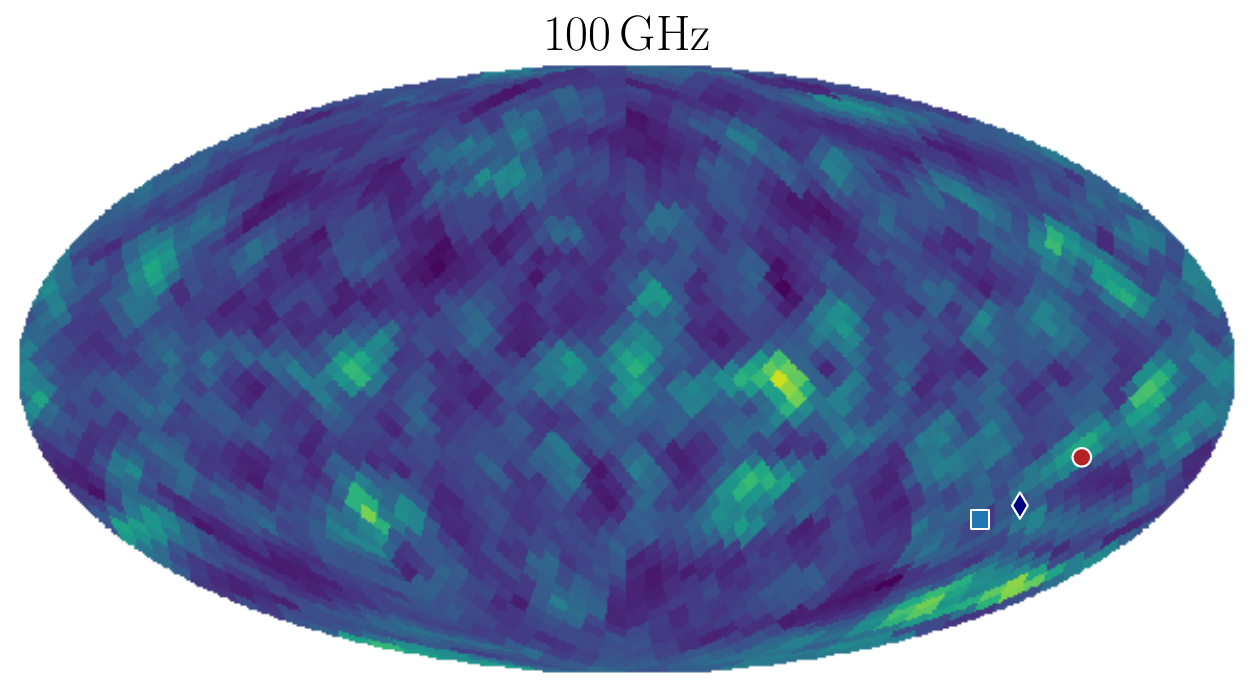}
    \includegraphics[width=0.45\linewidth]{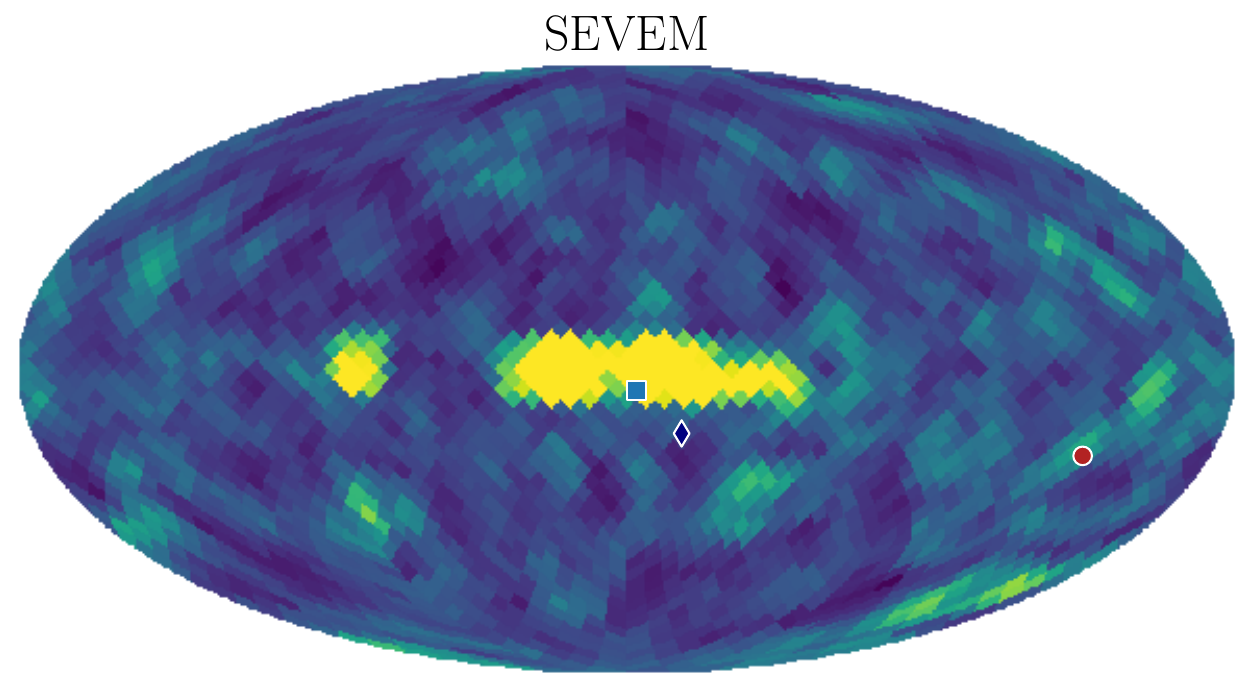}
    \includegraphics[width=0.45\linewidth]{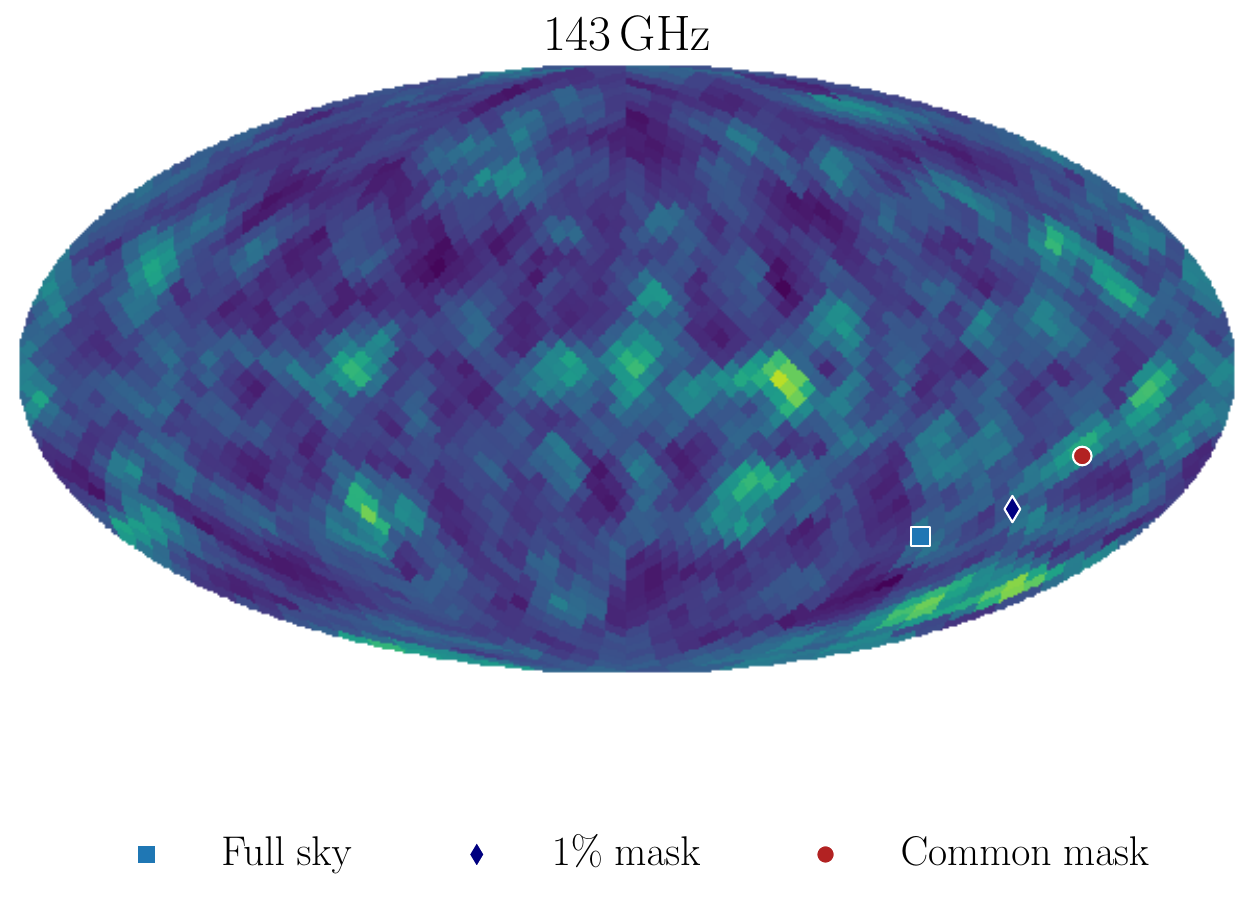}
    \includegraphics[width=0.45\linewidth]{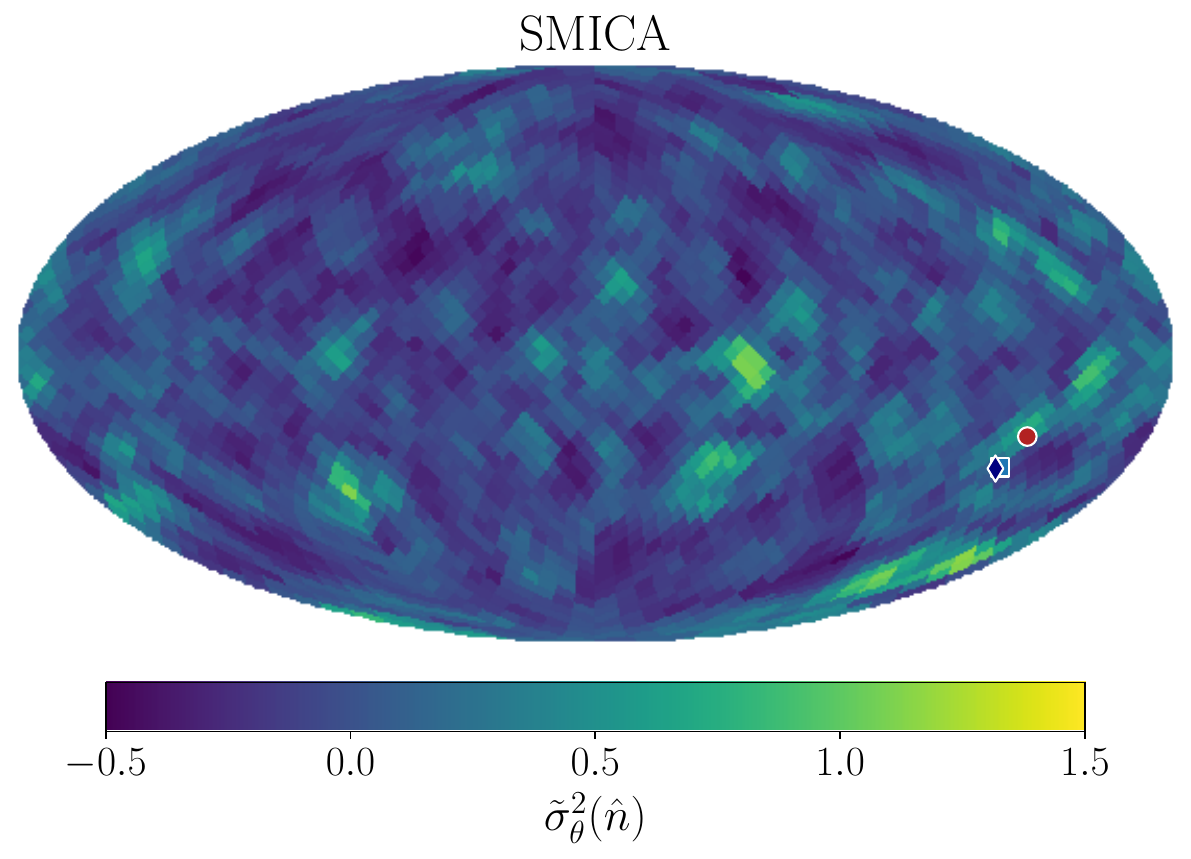}
    \caption{Normalized local-variance map, $\tilde{\sigma}_\theta^2(\hat{n})$, Eq.~\eqref{eq:ALV_normalized}, measured within $8^\circ$ disks applying the 1\% mask for all eight maps. The local-variance dipole directions are indicated as the markers for the three different sky cuts (indicated in the legend).}
    \label{fig:ALV_all}
\end{figure}

While Table~\ref{tab:p-values} shows the PTE of the anomaly statistics, Table~\ref{tab:stat_values} show the numerical values of the anomaly statistics for all maps and masks. Fig.~\ref{fig:Smu_all} shows the correlation functions, $C(\theta)$, and the PTE of the $S_\mu$-statistic as a function of the integration limit $\mu$ for all maps (Sec.~\ref{sec:S_12}). Fig.~\ref{fig:Rlmax_all} shows the low-$\ell$ power spectra and the PTE of the $R^{\ell_\max}$-statistic as a function of $\ell_\max$ for all maps (Sec.~\ref{sec:R27}). Fig.~\ref{fig:SQO_all} shows the quadrupole and octopole directions as traced by the MVs and the OAVs for the three sky cuts and for all maps (Sec.~\ref{sec:SQO}). Fig.~\ref{fig:ALV_all} shows the local-variance maps for the $1\%$-mask and the dipole direction of the local-variance maps for the three sky cuts and for all maps (Sec.~\ref{sec:ALV}). 

\begin{table*}[t]
    \hspace*{-1cm}
    \begin{tabular}{cl|cccc|cccc}
    \multicolumn{2}{c}{anomaly statistics} &70\,GHz &94\,GHz &100\,GHz &143\,GHz &\Commander &\NILC &\SEVEM &\SMICA\\
    \hline
                    &full sky: &7105 &7385 &6561 &6323 &5040 &6155 &4690 &5922 \\
    $S_{1/2}$       &1\% mask: &6519 &6424 &5965 &5688 &5346 &5781 &4060 &5742 \\
                    &com. mask: &1477 &1608 &1511 &1482 &1394 &1318 &1351 &1300 \\
    \hline
                    &full sky: &0.79 &0.79 &0.80 &0.80 &0.82 &0.79 &0.87 &0.80 \\
    $R^{27}$        &1\% mask: &0.79 &0.79 &0.79 &0.79 &0.79 &0.78 &0.83 &0.78 \\
                    &com. mask: &0.80 &0.80 &0.79 &0.79 &0.80 &0.80 &0.80 &0.80 \\
    \hline
                    &full sky: &0.80 &0.81 &0.81 &0.82 &0.76 &0.82 &0.72 &0.80 \\
    $S_\mathrm{QO}$ &1\% mask: &0.80 &0.81 &0.81 &0.81 &0.75 &0.82 &0.80 &0.80 \\
                    &com. mask: &0.60 &0.59 &0.61 &0.62 &0.63 &0.61 &0.57 &0.63 \\
    \hline
                    &full sky: &728 &738 &730 &735 &725 &718 &812 &715 \\
    $\sigma_{16}^2$ &3.6\% mask: &694 &702 &694 &692 &684 &700 &664 &696 \\
                    &com. mask: &581 &585 &587 &584 &615 &623 &576 &623 \\
    \hline
                    &full sky: &0.09 &0.21 &0.09 &0.08 &0.30 &0.10 &0.61 &0.10 \\
    $A_\mathrm{LV}$ &1\% mask: &0.09 &0.09 &0.09 &0.09 &0.11 &0.10 &0.19 &0.10 \\
                    &com. mask: &0.13 &0.13 &0.13 &0.13 &0.13 &0.14 &0.14 &0.14 \\
    \hline
    \end{tabular}
    \caption{Numerical values of the anomaly statistics considered in this work: Low correlation, $S_{1/2}$; parity asymmetry, $R^{27}$; quadrupole-octopole alignment, $S_\mathrm{QO}$; low northern variance at $N_\side=16$, $\sigma_{16}^2$; and local-variance asymmetry, $A_\mathrm{LV}$.}
    \label{tab:stat_values}
\end{table*}

\bibliography{main}
\bibliographystyle{aasjournal}

\end{document}